\documentclass[12pt]{article}
\usepackage{amssymb}
\usepackage{amsmath}
\usepackage{graphicx}
\usepackage{indentfirst}
\usepackage{cite}

\begin{document}

\title{Temperature-dependent cross sections \\
for meson-meson nonresonant reactions \\
in hadronic matter}
\author{Yi-Ping Zhang \and Xiao-Ming Xu
\and Hui-Jun Ge}
\date{}
\maketitle \vspace{-1cm}
\centerline{Department of Physics, Shanghai
University, Baoshan, Shanghai 200444, China}
\begin{abstract}
We present a potential of which the short-distance part is given by one gluon
exchange plus perturbative one- and two-loop corrections and of which the
large-distance part exhibits a temperature-dependent constant value. The
Schr\"odinger equation with this temperature-dependent potential yields a 
temperature dependence of the mesonic quark-antiquark relative-motion wave
function and of meson masses. The temperature dependence of the potential, 
the wave function and the meson masses brings about temperature dependence 
of cross sections for the nonresonant reactions $\pi \pi \to \rho \rho$ 
for $I=2$, $KK \to K^\ast K^\ast$ for $I=1$, $KK^\ast \to K^\ast K^\ast$ 
for $I=1$, $\pi K \to \rho K^\ast$ for $I=3/2$, $\pi K^\ast \to 
\rho K^\ast$ for $I=3/2$, $\rho K \to \rho K^\ast$ for 
$I=3/2$ and $\pi K^\ast \to \rho K$ for $I=3/2$.
As the temperature increases, the rise or fall of peak cross sections is 
determined by the increased radii of initial mesons, the loosened bound 
states of final mesons, and the total-mass difference of the initial and 
final mesons. The temperature-dependent cross sections and meson masses 
are parametrized.
\end{abstract}

\noindent
PACS: 25.75.-q; 13.75.Lb; 12.38.Mh

\noindent
Keywords: Meson-meson nonresonant reaction; Cross section; Quark-interchange
mechanism.

\newpage

\leftline{\bf 1. Introduction}
\vspace{0.5cm}
In spite of scarce experimental data of cross sections for inelastic 
meson-meson scattering, these cross sections can be calculated from
quark potential models, parton distributions or effective meson Lagrangians 
that respect various symmetries. This has been seen from the dissociation 
cross sections of $J/\psi$ in collisions with mesons, which have received 
much attention. For example, cross sections for reactions like $\pi+J/\psi 
\to D\bar {D}^* + D^*\bar {D} +D^\ast \bar {D}^\ast$ and $\rho+J/\psi \to 
D\bar {D}^* + D^*\bar {D}+D^\ast \bar {D}^\ast$ were obtained with the 
quark-interchange mechanism\cite{BS92,Swanson} in different quark 
potential models\cite{MBQ,WSB,BSWX,Xu02}. The quark-interchange models 
give the characteristic that cross sections for endothermic reactions 
first increase from threshold energies and then decrease as the 
center-of-mass energy of $J/\psi$ and hadron increases. A very small 
low-energy nucleon-$J/\psi$ break-up cross section was obtained by 
using the operator product expansion for the elastic scattering 
amplitude of $J/\psi$ \cite{SK,Peskin,BP}. The dissociation cross 
section for $\pi - J/\psi$, evaluated in QCD sum rules \cite{Duraes} in 
the soft-pion limit, increases gradually in contrast to rapid growth 
near the threshold energy obtained in meson exchange\cite{Mat98,Lin00,
Hag99,Hag01,Oh01} and quark interchange models \cite{MBQ,WSB,BSWX,Xu02}.

In comparison to meson-$J/\psi$ reactions, however, cross sections for 
inelastic scattering of a light meson by another light meson are rarely 
studied. Inelastic scattering between 
light mesons does occur in hadronic matter and the cross sections 
involved influence the time dependence of meson momentum distributions 
and flavor dependence of the measured momentum distributions at kinetic 
freeze-out. In order to understand this influence, the cross sections for
inelastic scattering and their relevant characteristics must be studied.
Since hadronic matter produced in Au-Au collisions at the Relativistic 
Heavy Ion Collider mainly consists of pions, rhos and kaons
\cite{adler,arsene,back,adams1,adcox,adams2,bearden1,bearden2},
in this work we pay 
attention only to the nonresonant reactions of the four mesons $\pi$, 
$\rho$, $K$ and $K^\ast$, which are taken to be governed by 
quark-interchange processes. In Ref. \cite{LX} we have calculated 
in-vacuum cross sections for the nonresonant reactions $\pi \pi \to 
\rho \rho$ for $I=2$, $KK \to K^\ast K^\ast$ for $I=1$, $KK^\ast \to 
K^\ast K^\ast$ for $I=1$, $\pi K \to \rho K^\ast$ for $I=3/2$,
$\pi K^\ast \to \rho K^\ast$ for $I=3/2$, $\rho K \to \rho 
K^\ast$ for $I=3/2$ and $\pi K^\ast \to \rho K$ for 
$I=3/2$. The cross sections for the seven endothermic 
reactions depend on the center-of-mass energy $\sqrt s$ of the two 
initial mesons and the energy where the maximum of cross section 
occurs is mainly determined by the maximum of $
\mid \vec {P}^\prime \mid / s\mid \vec {P} \mid$, where $\vec P$ 
and $\vec {P}^\prime$ are the momenta of the initial and final 
mesons in the center-of-mass frame, respectively. The endothermic 
reactions have 
maximum cross sections ranging from 0.47 mb to 1.41 mb. These cross 
sections are obtained with a quark-quark potential that includes
the linear confinement and one-gluon-exchange potential plus
perturbative one- and two-loop corrections \cite{Xu02}. At nonzero 
temperature the linear confinement is modified to become weaker. At 
a temperature below the critical temperature $T_c$ of the QCD phase 
transition, medium effects show up in the region where the 
quark-antiquark distance is larger than 0.3 fm \cite{KLP}. When the 
quark-antiquark distance is large enough, the quark-antiquark
potential at a given temperature becomes constant. 
This temperature-dependent 
potential must change the wave function of quark-antiquark 
relative motion in a meson and cross sections for inelastic meson-meson 
scattering are expected to change with temperature as well. Therefore, 
in this work we study the dependence of cross sections on the
temperature for the nonresonant reactions $\pi \pi \to \rho \rho$ for 
$I=2$, $KK \to K^\ast K^\ast$ for $I=1$, $KK^\ast \to K^\ast K^\ast$ 
for $I=1$, $\pi K \to \rho K^\ast$ for $I=3/2$, $\pi K^\ast 
\to \rho K^\ast$ for $I=3/2$, $\rho K \to \rho K^\ast$ for 
$I=3/2$ and $\pi K^\ast \to \rho K$ for $I=3/2$.
Until now the temperature dependence of the cross sections has not 
been studied in experiments or theory.

In the next section, we present a potential of which the 
short-distance part is given by perturbative QCD and the 
large-distance part is displayed in the lattice gauge results of 
Ref. \cite{KLP}. The Schr\"odinger equation with the potential is
solved to get the quark-antiquark relative-motion wave function of a 
meson. A convenient framework for application of the 
temperature-dependent potential to inelastic meson-meson scattering 
is the Born approximation for the quark-interchange processes.
In Section 3 we review formulas of cross sections for meson-meson
nonresonant reactions that are based on the Born approximation.
In Section 4 we show numerical results of cross sections and give 
relevant discussions. Parametrizations of the cross
sections are given. Conclusions are in the last section.

\vspace{0.5cm}
\leftline{\bf 2. Potential and wave functions of quark-antiquark 
relative motion}
\vspace{0.5cm}

In Ref. \cite{KLP} the heavy quark potential was assumed to be equal to the 
free energy of heavy quark-antiquark pair and the lattice calculations
provided the temperature dependence of the potential. 
When the distance $r$ between quark and antiquark 
is large enough, the potential at a given temperature exhibits a constant 
value and becomes a plateau. This value depends on the temperature $T$ 
and decreases with increasing temperature. The constant value at large 
distances can be parametrized as
\begin{equation}
V_{ab}(\vec {r})=-\frac {\vec {\lambda}_a}{2} \cdot \frac {\vec {\lambda}_b}{2}
 \frac {3}{4} D \left[ 1.3- \left( \frac {T}{T_c} \right)^4 \right] ,
\end{equation}
with $D=0.7$ GeV and $T_c=0.175$ GeV. Here $\vec \lambda$ are the 
Gell-Mann matrices for the color generators. $D$ is a fit parameter, but 
happens to equal the height of the plateau of $T=0.74T_c$.

The relative color orientation of two constituents (quarks and antiquarks here)
can be indicated by wave functions of the two constituents. The expectation 
value of
$\frac {\vec {\lambda}_a}{2} \cdot \frac {\vec {\lambda}_b}{2}$
depends on the relative color orientation, so does the heavy quark
potential given in Eq. (1). The dependence 
of the potential on the relative color orientation is supported by
hadronic physics and lattice calculations. We discuss evidence
in the three cases: (a) $T=0$, (b) $T>T_c$ and (c) $0<T<T_c$.

(a) Since the establishment of QCD the dependence of confinement with
$\frac {\vec {\lambda}_a}{2} \cdot \frac {\vec {\lambda}_b}{2}$ on
the relative color orientation has been widely used in hadronic
physics, for instance, meson-baryon scattering \cite{BDPK} and meson-meson 
scattering \cite{BSWX}. The expectation value of
$\frac {\vec {\lambda}_a}{2} \cdot \frac {\vec {\lambda}_b}{2}$ is
negative in the color singlet state and positive in any color octet
state. This complies with the fact that the color singlet state is
observed and the color octet states are not observed.

(b) When temperature is above $T_c$, it is 
expected in Ref. \cite{DHKK} that the interaction in an antitriplet
diquark system is only half as strong as in a color-singlet 
quark-antiquark system, i.e. obeys Casimir scaling, and the 
interaction depends on the relative color orientation.

(c) When temperature is between 0 and $T_c$, the situation is 
complicated. From Figs. 15, 17 and 20 in Ref. \cite{GHK}, we can derive that 
the color-singlet free energy does not equal the color-octet free energy at 
large distances in the region $0.907T_c \leq T<T_c$. It can also be expected 
from Fig. 3 of Ref. \cite{DHKK} that
the interaction in an antitriplet diquark system is nearly half as
strong as in a color-singlet quark-antiquark system, i.e. the interaction 
depends on the relative color orientation in the region $0.87T_c \leq T<T_c$.

Much attention has been paid to construction of the color octet
potential from perturbative QCD and lattice data in the study of hybrid mesons.
The color octet potential relies on the spatial extension of gluon 
field confined in hybrid mesons. In the string description the
symmetry of the gluon field is labeled by the spins $\Sigma$, $\Pi$
and $\Delta$ about the axis connecting the quark and antiquark,
a $PC$ value and an additional reflection symmetry for $\Sigma$ 
states. It was obtained in vacuum that the color octet potential is different
with respect to different symmetries of the gluon field \cite{Michael,BaPi}.
Temperature dependence of the gluon field affects temperature dependence of the
color octet potential. How does the gluon field confined in a meson depend on 
temperature at $0<T<T_c$? This is a difficult problem.

In summary, we still lack of the lattice study of the color octet potential at 
$0<T<T_c$ and the heavy quark potential depends on the relative color 
orientation. 

The large-distance plateau of the heavy quark potential indicates the onset of
string breaking at a certain distance \cite{DHKK}. 
The string breaking occurs when the energy stored in the string exceeds the 
mass of a light quark-antiquark pair \cite{Sjostrand}.
The combination of the light quark and the heavy antiquark and the combination
of the light antiquark and the heavy quark
form open heavy flavors. When the temperature is higher, light mesons in medium
more effectively take a kind of flip-flop recoupling of quark constituents 
\cite{Miyazawa,GS,Satz06hep} and string breaking occurs at a shorter distance. 
Hence, the plateau appears at a shorter distance and confinement gets weaker. 
Consequently, the wave function of the heavy quark-antiquark pair becomes wider
in space.

However, uncertainty of the heavy quark potential exists because
the potential is derived from Polyakov loop correlation functions. The 
Polyakov loop at the space coordinate $\vec {r}_1$ is indicated by 
$L(\vec {r}_1)$. The Polyakov loop correlation functions are related to the
free energy of heavy quark-antiquark pair $F(T,r)$
\begin{equation}
-T\ln <L(\vec {r}_1)L^+(\vec {r}_2)>=F(T,r)+C
\end{equation}
where $r=\mid \vec {r}_1 - \vec {r}_2 \mid$ and $C$ is a normalization 
constant. The free energy leads to the internal energy by
\begin{equation}
U(T,r)=F(T,r)+TS(T,r)
\end{equation}
The entropy $S(T,r)=-\partial F(T,r) / \partial T$ is independent of $r$ at 
large distances and depends on $r$ at the other distances. If $TS$ is small,
$U(T,r) \approx F(T,r)$; otherwise, the internal energy deviates from the
free energy. Taking the free energy as the heavy quark potential like Ref. 
\cite{KLP} is an approximation. Then the internal energy was suggested in Refs.
\cite{KKPZ,ZKKP} as the heavy quark potential.
Very recently the heavy quark potential is argued to be the internal energy 
which is the expectation value of the difference of the Hamiltonians
with and without the heavy quark-antiquark pair at rest
\cite{Satz09,KS}, and this results in dissociation temperatures of heavy 
quarkonia that agree quite well with the values from lattice studies.

How to find the correct heavy quark potential is an important
problem but has not been solved so far. 
It has been proposed that the heavy quark potential takes the form 
$\xi U(T,r)+(1-\xi )F(T,r)$ where the quantity $\xi$ is between 0 and 1 and was
determined in models \cite{DPS1,SZ,Wong05,Wong,Wong07,Alberico,Satz06}. 
Since the present lattice calculations provide the internal energy
$U$ that includes the internal energy of the heavy quark-antiquark
pair and the gluon internal energy difference $U_{\rm g}(T,r)
-U_{{\rm g}0}(T)$ where $U_{\rm g}(T,r)$ and $U_{{\rm g}0}(T)$ 
correspond to the gluon internal energies in the presence and absence of 
the heavy quark-antiquark pair, respectively, it was proposed by Wong 
\cite{Wong05} that 
the heavy quark potential should be $U(T,r)-[U_{\rm g}(T,r)
-U_{{\rm g}0}(T)]$. In the local energy-density approximation that adopts 
an equation of state for the quark-gluon plasma, the heavy quark potential 
can be represented as a linear combination of $F$ and $U$. With $F$ and $U$
obtained in lattice calculations in quenched QCD, the potential
gives spontaneous dissociation temperatures of heavy quarkonia that
agree with those obtained from spectral analyses in quenched QCD \cite{Wong07}.
This is an
improvement in comparison to the use of $F$ as a potential. The work of 
Wong implies that the heavy quark potential properly defined gives reliable 
results. Even though the method of Wong has not been applied to hadronic 
matter, we still expect that the heavy quark potential for hadronic matter, if
obtained, can produce more reliable cross sections for the meson-meson
nonresonant reactions in the nonzero temperature region of hadronic matter
than those from $F$.

At very small distances $r<0.01$ fm, the potential obtained from
one-gluon exchange plus one- and two-loop corrections in 
perturbative QCD is \cite{BT}
\begin{equation}
V_{ab}(\vec {r}) =
\frac {\vec {\lambda}_a}{2} \cdot \frac {\vec {\lambda}_b}{2}
\frac {12\pi}{25rw} \left [ 1+\left( 2\gamma_E+\frac {53}{75}\right)
\frac {1}{w} -\frac {462}{625} \frac {\ln w}{w} \right ] ,
\end{equation}
where $\gamma_E$ is the Euler's constant and $w=\ln(1/\Lambda^2_{\overline
{\rm MS}}r^2)$ with the QCD scale parameter $\Lambda_{\overline {\rm MS}}$
determined by Eq. (2.13) in Ref. \cite{BT}.

An interpolation between the constant confinement at large distances given 
in Eq. (1) and the spin-independent perturbative potential given in Eq. (4) 
produces a central spin-independent potential
\begin{equation}
V_{ab}(\vec {r})=-\frac {\vec {\lambda}_a}{2} \cdot \frac {\vec {\lambda}_b}{2}
\frac {3}{4} D \left[ 1.3- \left( \frac {T}{T_c} \right)^4 \right]
\tanh (Ar) + \frac {\vec {\lambda}_a}{2} \cdot \frac {\vec {\lambda}_b}{2}
\frac {6\pi}{25} \frac {v(\lambda r)}{r} \exp (-Er)  ,
\end{equation}
where
\begin{equation}
A=1.5[0.75+0.25(\frac {T}{T_c})^{10}]^6~{\rm GeV},
\end{equation}
and
\begin{equation}
E=0.6~{\rm GeV},
\end{equation}
are fit parameters, and
\begin{equation}
\lambda=\sqrt {3b_0/16\pi^2\alpha'},
\end{equation}
$\alpha'=1.04~{\rm GeV}^{-2}$ is the Regge slope and $b_0=11-\frac{2}{3} N_f$ 
with the quark flavor number $N_f=4$ \cite{BT}. 
This potential is different from the 
parametrizations given by Digal et al. \cite{DPS1} and Wong \cite{Wong02}.
The potential contains the dimensionless function \cite{BT}
\begin{equation}
v(x)=\frac
{4b_0}{\pi} \int^\infty_0 \frac {dQ}{Q} (\rho (\vec {Q}^2) -\frac {K}{\vec
{Q}^2}) \sin (\frac {Q}{\lambda}x) ,
\end{equation}
with $K= 3/16\pi^2\alpha'$, where $Q$ is the absolute value of gluon momentum
$\vec Q$ and $\rho (\vec {Q}^2)$ is given by Buchm\"uller and Tye \cite{BT}.
The quantity $\rho - \frac {K}{\vec {Q}^2}$ arises from one-gluon exchange
and perturbative one- and two-loop corrections. $\exp (-Er)$ is a medium 
modification factor to the potential of one-gluon exchange plus perturbative
one- and two-loop corrections. The temperature correction to the
one-gluon-exchange potential with the limit shown in Eq. (4) is the difference
between the second term in Eq. (5) and the perturbative potential
$\frac {\vec {\lambda}_a}{2} \cdot \frac {\vec 
{\lambda}_b}{2} \frac {6\pi}{25} \frac {v(\lambda r)}{r}$ \cite{BT}.
The temperature correction is completely negligible at very short distances 
and obvious at intermediate and large distances.

The parametrization in
Eq. (5) versus lattice gauge results is plotted in Fig. 1. It is clearly
seen that a lower plateau at large distances corresponds to a higher
temperature. Plateaus at $T/T_c=0.97$, 0.94, 0.9, 0.84 approximately begin at
$r=1.15$ fm, 1.18 fm, 1.26 fm, 1.38 fm, respectively. Hence, a higher plateau
begins at a larger distance. Confinement can be assumed to be 
flavor-independent in hadronic physics. For example, it was shown in Refs.
\cite{DGG,IS1,IS2,IS3,GI,CI} 
that quark-quark potentials with flavor-independent confinement, a 
Coulomb term and hyperfine interactions can consistently describe a large body 
of data like masses from light to heavy hadrons. The light hadrons may consist 
of only up and down quarks and the heavy hadrons may contain charm and bottom
quarks. The flavor dependence of the quark-quark potentials is relevant to 
quark masses in the hyperfine interactions. 
The success of the potentials renders that the flavor independence of
confinement is universal and the hyperfine interactions must be 
flavor-dependent. Therefore, the potential in Eq. (5) is reasonably
flavor-independent. The potential obtained by Karsch et al. for heavy
quarks is applied to light quarks. It is shown by the lattice calculations that
screening sets in at distances $r \approx 0.3$ fm. However, present lattice
calculations at finite temperatures are probably not yet precise enough to
reach sufficiently short distances. 
Hence, a lattice-based potential has a degree of freedom
in choosing its form at short distances. We construct the potential in Eq. (5)
from the fact that the quark-quark interaction at $r<0.01$ fm originates from 
one-gluon exchange plus loop corrections in perturbative QCD. The degree of 
freedom is removed at $r<0.01$ fm.

The medium screening obtained in the lattice gauge calculations
at present affects only the central spin-independent potential. This allows us
to keep using the spin-spin interaction relevant to perturbative QCD, as done
by Wong \cite{Wong02}. In our work, the spin-spin interaction arises not only 
from one-gluon exchange but also from
perturbative one- and two-loop corrections 
\cite{Xu02}
\begin{equation}
V_{\rm ss}=
- \frac {\vec {\lambda}_a}{2} \cdot \frac {\vec {\lambda}_b}{2}
\frac {16\pi^2}{25} \delta^3(\vec {r})  \frac {\vec {s}_a \cdot \vec
{s}_b}{m_am_b}
+ \frac {\vec {\lambda}_a}{2} \cdot \frac {\vec {\lambda}_b}{2}
  \frac {4\pi}{25} \frac {1}{r}
\frac {d^2v(\lambda r)}{dr^2} \frac {\vec {s}_a \cdot \vec {s}_b}{m_am_b} ,
\end{equation}
where $\vec {s}_a$ and $m_a$ are the spin and mass of constituent
quarks or antiquarks labeled as $a$, respectively. This expression of 
the spin-spin interaction comes from the application of Eq. (7e) in the 
transformed Hamiltonian obtained by Chraplyvy \cite{Chra} from the
two-constituent Hamiltonian that includes the relativistic potential
originating from one-gluon exchange plus perturbative one- and two-loop
corrections \cite{Xu02}. The transformed Hamiltonian was obtained from 
an application of the Foldy-Wouthuysen canonical transformation to a 
relativistic two-particle Hamiltonian. The spin-spin interaction is 
related to the terms of the direct product of two Dirac $\alpha$ 
matrices \cite{BD} in the relativistic potential.

Given the masses $m_{\rm u}=m_{\rm d}=0.32$ GeV for the up and down 
quarks, the Schr$\rm \ddot o$dinger equation with the central 
spin-independent potential in Eq. (5) is solved at $T=0$ to obtain a 
radial wave function $R_{q\bar {q}}(T=0,r)$ for the quark-antiquark 
relative motion of $\pi$ and $\rho$ mesons. Assuming all the mesons 
in the ground-state pseudoscalar octet and the ground-state vector 
nonet taking the same spatial wave function of quark-antiquark 
relative motion as the $\pi$ and $\rho$ mesons, the spin-spin 
interaction leads to the mass splitting between a pseudoscalar 
meson and a vector meson with the same isospin quantum number
\begin{equation}
<V_{\rm ss}>=\frac {16\pi}{75m_am_b}[R^2_{q\bar {q}}(T=0,r=0)
-\int^\infty_0dr r \frac
{d^2v(\lambda r)}{dr^2}R^2_{q\bar {q}}(T=0,r)] ,
\end{equation}
where $R_{q\bar {q}}$ satisfies the normalization condition
$\int^\infty_0 dr r^2 R^2_{q\bar {q}}(T,r)=1$.
At $m_{\rm s}=0.5$ GeV for the strange-quark mass,
the spin-spin interaction yields the
mass splittings $m_\rho -m_\pi=0.5989$ GeV, $m_{K^\ast}-m_K=0.3833$ GeV 
and $\frac {1}{3}m_{\omega}+\frac {2}{3}m_{\phi}-m_\eta=0.3622~{\rm GeV},$ 
where $m_i$ $(i=\pi, \rho, K, K^\ast, \eta, \omega, \phi)$ represent the 
masses of $\pi, \rho, K, K^\ast, \eta, \omega$ and $\phi$, respectively.
These mass splittings can be compared to the experimental values 0.6304 GeV,
0.3963 GeV and 0.3930 GeV, respectively.

The spin-averaged mass of a spin-0 meson and a spin-1 meson with the same 
isospin is one-fourth of the spin-0 meson mass plus three-fourths of the 
spin-1 meson mass \cite{MBQ}. In Table 1 we list vacuum masses of $\pi$, 
$\rho$, $K$ and $K^\ast$, the spin-averaged mass $\overline {m}_{\pi \rho}$
of $\pi$ and $\rho$ and the one $\overline {m}_{K K^\ast}$ of $K$ and $K^\ast$.
The theoretical values of 
$\overline {m}_{\pi \rho}$, $\overline {m}_{K K^\ast}$, $m_\rho$, $m_K$ and
$m_{K^*}$ approach the corresponding experimental data. The pion 
mass from our calculations almost doubles the experimental value. These are
understandable. The potential given in Eq. (5) has the behavior of $\tanh (Ar)$
at large distances and cannot mimic the linear confinement. In vacuum
the pion with the
lightest mass among mesons has the largest radius and is sensitive to the 
potential behavior at large distances. Therefore, the experimental datum of 
pion mass must differ from the value derived from the potential. For $\rho$, 
$K$ and $K^\ast$ with masses larger than $\pi$, less sensitivity to the 
potential
behavior at large distances does not lead to a large separation of the
experimental and theoretical masses. If the theoretical pion mass is used,
cross sections for pions in collisions with mesons are not reliable. Hence, the
experimental values of meson masses are used in calculating meson-meson cross
sections at $T=0$ GeV in this work and, as will see, the cross sections are 
similar to those obtained from the potential with the linear confinement in
Ref. \cite{LX}.

If the quark masses are set equal to zero, the QCD Hamiltonian is
symmetric under the chiral group $SU(3) \times SU(3)$. Although
the spontaneous breakdown of the chiral symmetry is widely expected, the
problem of calculating the spontaneous chiral symmetry  breaking in
QCD has not yet been satisfactorily solved \cite{Weinberg,Pokorski}. 
However, the spontaneous
symmetry breaking can be related to the quark condensate which
vanishes in perturbative QCD and does not equal zero in the
non-perturbative region. With the
explicit breakdown of the chiral symmetry due to the current quark
masses, the square of pion mass is proportional to the product of the
quark condensate and the sum of the up and down quark masses at the
lowest order in chiral perturbation theory. Therefore, the small pion
mass comes from the small quark masses and the non-vanishing quark
condensate in the non-perturbative region where confinement sets in.
While only the spontaneous symmetry breaking occurs, Goldstone bosons
and quarks are massless and the soft-pion theorems obtained in the
vanishing pion four-momentum are exact \cite{FGH}. While the explicit chiral 
symmetry breaking also occurs, bosons get masses and 
the soft-pion theorems become approximate. The
deviation of the theorems from the experimental data has been studied in
chiral perturbation theory and the soft-pion theorems have been corrected 
\cite{GL1,GL2,BCEGS,CGL}. In our work all the mesons in the ground-state
pseudoscalar octet and the ground-state vector nonet are assumed to take
the same spatial wave function of quark-antiquark relative
motion as the pion and rho mesons, but the quark masses are the
constituent masses which are much larger than the current quark
masses used in chiral perturbation theory. The constituent quark masses show
the explicit chiral symmetry breaking so that mesons get masses
and the soft-pion theorems are violated. The cross sections that we will 
obtain from the constituent quark masses and the
nonzero meson masses must deviate from the cross sections obtained
from vanishing masses of quarks and mesons. Therefore, the deviation of our 
results from the soft-pion theorems should be expected.

\vspace{0.5cm}
\leftline{\bf 3. Formulas for cross sections}
\vspace{0.5cm}

Let $m_i$ and $P_i = (E_i, \vec {P}_i)$ be the mass and the four-momentum
of meson $i$ ($i=q_1\bar {q}_1, q_2\bar {q}_2, q_1\bar {q}_2, q_2\bar {q}_1$),
respectively. The Mandelstam variables for the reaction
$A(q_{1}\overline{q}_{1})+B(q_{2}\overline{q}_{2})\to
C(q_{1}\overline{q}_{2})+D(q_{2}\overline{q}_{1})$ are
$s=(E_{q_1\bar {q}_1}+E_{q_2\bar {q}_2})^2
-(\vec {P}_{q_1\bar {q}_1}+\vec {P}_{q_2\bar {q}_2})^2$ and
$t=(E_{q_1\bar {q}_1}-E_{q_1\bar {q}_2})^2
-(\vec {P}_{q_1\bar {q}_1}-\vec {P}_{q_1\bar {q}_2})^2$. In the
center-of-mass frame the meson $A(q_{1}\overline{q}_{1})$ has the 
momentum $\vec {P} = \vec {P}_{q_1\bar {q}_1}= -\vec {P}_{q_2\bar {q}_2}$
and the meson $C(q_{1}\overline{q}_{2})$ has the momentum
$\vec {P}^\prime = \vec {P}_{q_1\bar {q}_2}= -\vec {P}_{q_2\bar {q}_1}$.
$\vec {P}$ and $\vec {P}^\prime$ are expressed in terms of $s$ by
\begin{equation}
|\vec {P}(\sqrt{s})|^{2}=\frac{1}{4s}\left\{ \left[ s-\left(
m_{q_1\bar {q}_1}^{2}+m_{q_2\bar {q}_2}^{2}\right) \right]^{2}
-4m_{q_1\bar {q}_1}^{2}m_{q_2\bar {q}_2}^{2} \right\},
\end{equation}
\begin{equation}
|\vec {P}^{\prime}(\sqrt{s})|^{2}=\frac{1}{4s}\left\{ \left[
s-\left( m_{q_1\bar {q}_2}^2+m_{q_2\bar {q}_1}^2\right)\right]^2
-4m_{q_1\bar {q}_2}^2m_{q_2\bar {q}_1}^2
\right\} .
\end{equation}%
Denote the angle between $\vec {P}$ and $\vec {P}^\prime$ by $\theta$. The
cross section for
$A(q_{1}\overline{q}_{1})+B(q_{2}\overline{q}_{2})\to
C(q_{1}\overline{q}_{2})+D(q_{2}\overline{q}_{1})$ is
\begin{equation}
\sigma =\frac{1}{32\pi s}\frac{|\vec{P}^{\prime }(\sqrt{s})|
}{|\vec{P}(\sqrt{s})|}\int_{0}^{\pi }d\theta
|\mathcal{M}_{\rm fi} (s,t)|^{2}\sin \theta .
\end{equation}
This formula provides the dependence on the total energy $\sqrt s$ of 
the two initial mesons in the center-of-mass frame and is valid 
for the interchange of the two quarks ($q_1$ and $q_2$) or of the two 
antiquarks ($\bar {q}_1$ and $\bar {q}_2$). The interchange of quarks 
brings about two forms of scattering that may lead to 
different values of the transition amplitude ${\cal M}_{\rm fi}$.
The forms are known as the prior form and the post form \cite{MM,BBS,WC}. 
Scattering in the prior form means that gluon exchange takes place 
prior to the quark or antiquark interchange. The transition amplitude in 
the prior form is \cite{LX}
\begin{eqnarray}
{\cal M}_{\rm fi}^{\rm prior} & = &
\sqrt {2E_{q_1\bar {q}_1}2E_{q_2\bar {q}_2}2E_{q_1\bar {q}_2}
2E_{q_2\bar {q}_1}}
\int \frac {d^3 p_{q_1\bar {q}_2}}{(2\pi)^3}
     \frac {d^3 p_{q_2\bar {q}_1}}{(2\pi)^3}      \nonumber    \\
& & \psi^+_{q_1\bar {q}_2} (\vec {p}_{q_1\bar {q}_2})
\psi^+_{q_2\bar {q}_1} (\vec {p}_{q_2\bar {q}_1})
(V_{q_1\bar {q}_2}+V_{\bar {q}_1 q_2}+V_{q_1 q_2}+V_{\bar {q}_1 \bar {q}_2})
\psi_{q_1\bar {q}_1} (\vec {p}_{q_1\bar {q}_1})
\psi_{q_2\bar {q}_2} (\vec {p}_{q_2\bar {q}_2}) ,   \nonumber   \\
\end{eqnarray}
where $\psi_{ab} (\vec {p}_{ab})$ is the product of color, spin, flavor and
momentum-space wave functions of the relative motion of constituents $a$ and
$b$ and satisfies $\int \frac {d^3p_{ab}}{(2\pi)^3} \psi^+_{ab} (\vec {p}_{ab})
\psi_{ab} (\vec {p}_{ab}) =1$. The relative momentum of $a$ and $b$ is 
$\vec {p}_{ab}$. Scattering in the post form means that the quark or antiquark
interchange is followed by gluon exchange. The transition amplitude in the
post form is \cite{LX}
\begin{eqnarray}
{\cal M}_{\rm fi}^{\rm post} & = &
\sqrt {2E_{q_1\bar {q}_1}2E_{q_2\bar {q}_2}2E_{q_1\bar {q}_2}
2E_{q_2\bar {q}_1}}        \nonumber   \\
& & \left( \int \frac {d^3 p_{q_1\bar {q}_1}}{(2\pi)^3}
     \frac {d^3 p_{q_1\bar {q}_2}}{(2\pi)^3}
\psi^+_{q_1\bar {q}_2} (\vec {p}_{q_1\bar {q}_2})
\psi^+_{q_2\bar {q}_1} (\vec {p}_{q_2\bar {q}_1})
V_{q_1\bar {q}_1}
\psi_{q_1\bar {q}_1} (\vec {p}_{q_1\bar {q}_1})
\psi_{q_2\bar {q}_2} (\vec {p}_{q_2\bar {q}_2})   \right.
           \nonumber   \\
& & + \int \frac {d^3 p_{q_2\bar {q}_2}}{(2\pi)^3}
           \frac {d^3 p_{q_2\bar {q}_1}}{(2\pi)^3}
\psi^+_{q_1\bar {q}_2} (\vec {p}_{q_1\bar {q}_2})
\psi^+_{q_2\bar {q}_1} (\vec {p}_{q_2\bar {q}_1})
V_{\bar {q}_2q_2}
\psi_{q_1\bar {q}_1} (\vec {p}_{q_1\bar {q}_1})
\psi_{q_2\bar {q}_2} (\vec {p}_{q_2\bar {q}_2})       \nonumber   \\
& & + \int \frac {d^3 p_{q_1\bar {q}_2}}{(2\pi)^3}
           \frac {d^3 p_{q_2\bar {q}_1}}{(2\pi)^3}
\psi^+_{q_1\bar {q}_2} (\vec {p}_{q_1\bar {q}_2})
\psi^+_{q_2\bar {q}_1} (\vec {p}_{q_2\bar {q}_1})
V_{q_1 q_2}
\psi_{q_1\bar {q}_1} (\vec {p}_{q_1\bar {q}_1})
\psi_{q_2\bar {q}_2} (\vec {p}_{q_2\bar {q}_2})       \nonumber   \\
& & \left. + \int \frac {d^3 p_{q_1\bar {q}_2}}{(2\pi)^3}
           \frac {d^3 p_{q_2\bar {q}_1}}{(2\pi)^3}
\psi^+_{q_1\bar {q}_2} (\vec {p}_{q_1\bar {q}_2})
\psi^+_{q_2\bar {q}_1} (\vec {p}_{q_2\bar {q}_1})
V_{\bar {q}_1 \bar {q}_2}
\psi_{q_1\bar {q}_1} (\vec {p}_{q_1\bar {q}_1})
\psi_{q_2\bar {q}_2} (\vec {p}_{q_2\bar {q}_2})  \right) .    \nonumber   \\
\end{eqnarray}

The transition amplitudes in the prior form and in the post form are 
equal to one another when the potential and the wave function of 
quark-antiquark relative motion are those used in the 
Schr\"odinger equation \cite{MM,BBS,WC}. Otherwise, 
${\cal M}_{\rm fi}^{\rm prior} \not= {\cal M}_{\rm fi}^{\rm post}$
for inelastic scattering. The inequality yields different cross sections
corresponding to the two forms, which is the so-called post-prior 
discrepancy\cite{MM,BBS,WC}.

In the Schr\"odinger equation we have only used the central spin-independent
potential in Eq. (5) since the spin-spin interaction in Eq. (10) contains the
delta function that can not be correctly dealt with in the equation. 
But in the transition amplitudes where the Fourier transform of
the spin-spin interaction can be correctly dealt with, we use the 
Fourier transform of both the central spin-independent 
potential and the spin-spin interaction:
\begin{eqnarray}
V_{ab}\left( \vec {Q}\right) &=& -\frac{ \vec {\lambda }_{a}}{2}
\cdot \frac{\vec {\lambda }_{b}}{2}\frac{3}{4} D
\left[ 1.3- \left( \frac {T}{T_c} \right)^4 \right]
\left[ (2\pi)^3\delta^3 (\vec {Q}) - \frac {8\pi}{Q}
\int^\infty_0 dr \frac {r\sin (Qr)}{\exp (2Ar)+1} \right]
                               \notag \\
& &
+\frac{ \vec {\lambda }_{a}}{2} \cdot \frac{\vec {\lambda }_{b}}{2} 64 \pi E
\int^\infty_0 dq \frac {\rho (q^2) -\frac {K}{q^2}}{(E^2+Q^2+q^2)^2-4Q^2q^2}
                               \notag \\
& & -\frac{\vec {\lambda }_{a}}{2}
\cdot \frac{\vec {\lambda }_{b}}{2}\frac{16\pi ^{2}}{25}\frac{
\vec {s}_{a}\cdot \vec {s}_{b}}{m_{a}m_{b}}
+\frac{\vec {\lambda }_{a}}{2}\cdot \frac{\vec {\lambda }_{b}}{
2}\frac{16\pi ^{2}\lambda }{25Q}\int_{0}^{\infty}dx\frac{d^{2}v\left(
x\right) }{dx^{2}}\sin \left( \frac{Q}{\lambda }x\right)
\frac{\vec {s}_{a}\cdot \vec {s}_{b}}{m_{a}m_{b}} .
                                 \notag \\
\end{eqnarray}
Therefore, the post-prior discrepancy occurs in our calculations. 
We take the average of the cross section in the prior form and the 
one in the post form. Each of the two cross sections related to the
prior form and the post form is the unpolarized cross section 
obtained from the cross section in Eq. (14)
\begin{equation}
\sigma ^{\rm unpol} (\sqrt {s})=
\frac{1}{(2S_{A}+1)(2S_{B}+1)}\sum _{S}(2S+1)\sigma(S,m_S,\sqrt {s}) ,
\end{equation}
where $S_A$ and $S_B$ are the spins of $A$ and $B$, respectively, and $S$ is
the total spin of the two mesons allowed by the reaction $A+B \to C+D$. The 
cross section $\sigma ^{\rm unpol} (\sqrt {s})$ is independent of the 
magnetic projection quantum number $m_S$ of $S$.

\vspace{0.5cm}
\leftline {\bf 4. Numerical results and discussions}
\vspace{0.5cm}

The quark masses $m_{\rm u}=m_{\rm d}=0.32$ GeV and $m_{\rm s}=0.5$ GeV
determined in the fit to the experimental mass splittings in Section 2 are 
kept unchanged in hadronic matter. They are used in the Schr\"odinger
equation with the central spin-independent but temperature-dependent 
potential in Eq. (5). The lowest-energy $S$-wave
solution of the Schr\"odinger equation is a temperature-dependent
radial wave function of the quark-antiquark relative motion of mesons in the
ground-state pseudoscalar octet and the ground-state vector nonet. When 
the temperature increases, the peak of $r$ times the $S$-wave radial wave
function, $rR_{q\bar {q}}(T,r),$ moves to larger quark-antiquark distances 
and the meson's root-mean-square radius increases.
This reflects the phenomenon that with increasing temperature
any bound state becomes looser and looser
while confinement gets weaker, i.e. the potential plateau at large
distances decreases with increasing temperature and a higher plateau 
begins at a larger distance.

Another consequence of the temperature-dependent potential in Eq. (5)
is that masses of $\pi$, $\rho$, $K$ and $K^\ast$ decrease with  
increasing temperature in the region $0.6 \leq T/T_c \leq 0.99$.
The mass splitting of a spin-0 meson and a spin-1 meson with the same 
isospin is calculated according to
\begin{equation}
<V_{\rm ss}>=\frac {16\pi}{75m_am_b}[R^2_{q\bar {q}}(T,r=0)
-\int^\infty_0dr r \frac
{d^2v(\lambda r)}{dr^2}R^2_{q\bar {q}}(T,r)] .
\end{equation}
The spin-averaged mass of a spin-0 meson and a spin-1 meson with the same
isospin is one-fourth of the spin-0 meson mass plus three-fourths of the spin-1
meson mass \cite{MBQ}. The spin-averaged mass of $\pi$ and $\rho$ equals the
sum of quark mass, antiquark mass and the energy of the relative motion 
obtained from the Schr\"odinger equation.
Since for the quark-antiquark relative motion we take the same spatial wave
functions of $K$ and $K^\ast$ as that of $\pi$ and $\rho$ mesons, the 
spin-averaged mass of $K$ and $K^*$ equals the sum of quark mass, antiquark 
mass and the nonrelativistic-Hamiltonian
expectation value of the wave function. After mass 
splittings and spin-averaged masses are obtained, we find meson masses of 
which the mass of the spin-0 meson is the
spin-averaged mass minus three-fourths of the mass splitting of the spin-0
and spin-1 mesons, and the mass of the spin-1 meson is the spin-averaged mass
plus one-fourth of the mass splitting.
The temperature-dependent meson masses are plotted in Fig.
2. The reason for the falloff of masses with increasing temperature
is that the first term of the potential related to the large-distance 
plateau gives a smaller contribution at higher temperature. When 
$T \to T_c$, the masses of the $\pi$ and $\rho$ mesons approach 0 and 
0.006 GeV, respectively. This indicates that the $\pi$ and $\rho$ mesons 
are almost massless at a temperature very close to $T_c$. In addition to 
the falloff of masses, the mass splitting of $\pi$ and $\rho$ and the one 
of $K$ and $K^\ast$ are both close to zero at $T \to T_c$. When $T \to T_c$, 
the masses of $K$ and $K^\ast$ approach finite values 0.179 GeV and 0.183 GeV, 
respectively. The kaons and vector kaons become degenerate in mass near $T_c$. 
Similar to Ref. [3], a constant of 0.88857 GeV is subtracted from the energy of
the Schr\"odinger equation for $\pi$ and $\rho$ and from the 
nonrelativistic-Hamiltonian expectation value of the wave function for $K$ and
$K^\ast$. This subtraction makes the theoretical values of the spin-averaged
masses at $T=0$ approach the experimental data and the masses of $\pi$ and 
$\rho$ go to zero at $T \to T_c$. The subtraction does not influence the sizes
of meson bound states and the mass splittings.

The meson masses in units of GeV are parametrized as
\begin{equation}
m_{\pi}=0.41 \left[ 1-\left( \frac{T}{1.05T_c} \right)^{11.88} \right]^{3.81} ,
\end{equation}
\begin{equation}
m_{\rho}=0.7 \left[ 1-\left( \frac{T}{T_c} \right)^{4.29} \right]^{1.14} ,
\end{equation}
\begin{equation}
m_{K}=0.63 \left[ 1-\left( \frac{T}{1.15T_c} \right)^{9.31} \right]^{4.4} ,
\end{equation}
\begin{equation}
m_{K^\ast}=0.84 \left[ 1-\left( \frac{T}{1.05T_c} \right)^{4.16} \right] ,
\end{equation}
which are valid in the region $0.6 \leq T/T_c \leq 0.99$. Since the temperature
of hadronic matter is generally larger than 0.11 GeV and smaller than $T_c$,
the meson masses and cross sections shown in the next paragraph in the
temperature region are sufficient for studies concerned with hadronic matter.

The wave function of quark-antiquark relative motion of the $\pi$ and $\rho$
mesons is taken to be the same as those of the other ground-state mesons.
With $m_{\rm u}=m_{\rm d}=0.32$ GeV obtained in Section 2,
the experimental data of $S$-wave $I=2$ elastic phase shifts for $\pi \pi$
scattering in vacuum \cite{Col71,Dur73,Hoo77,Los74} are reproduced with the
potential in Eq. (17).
Keeping quark masses at the values determined in the fit to the experimental
mass splittings, the dependence of the transition
amplitude $\mathcal{M}_{\rm fi}$ on temperature comes from the wave function,
the potential in Eq. (17) and the meson masses. The transition amplitude 
contains also color, spin and flavor matrices that are not affected by the 
temperature. Since the meson masses depend on temperature, threshold energies 
for inelastic meson-meson scattering depend on temperature. The temperature
dependence of the potential, the quark-antiquark wave function and the
meson masses leads to temperature-dependent cross sections for inelastic
meson-meson scattering. Unpolarized cross sections for the seven reactions 
$\pi \pi \to \rho \rho$ for $I=2$, $KK \to K^\ast K^\ast$ for $I=1$, 
$KK^\ast \to K^\ast K^\ast$ for $I=1$, $\pi K \to \rho K^\ast$ for 
$I=3/2$, $\pi K^\ast \to \rho K^\ast$ for $I=3/2$,
$\rho K \to \rho K^\ast$ for $I=3/2$, and $\pi K^\ast \to \rho K$ 
for $I=3/2$ are plotted in Figs. 3-9. For convenient application 
the numerical cross sections are parametrized as
\begin{eqnarray}
\sigma^{\rm unpol} & = &
a_1 \left( \frac {\sqrt {s} -\sqrt {s_0}}{b_1} \right)^{c_1}
\exp \left[ c_1 \left( 1-\frac {\sqrt {s} -\sqrt {s_0}}{b_1} \right) \right]
                   \notag   \\
& & + a_2 \left( \frac {\sqrt {s} -\sqrt {s_0}}{b_2} \right)^{c_2}
\exp \left[ c_2 \left( 1-\frac {\sqrt {s} -\sqrt {s_0}}{b_2} \right) \right] .
\end{eqnarray}
Time-consuming computations determine values of the parameters $a_1$, $b_1$,
$c_1$, $a_2$, $b_2$ and $c_2,$ which are shown in Tables 2 and 3.

At the threshold energy of an endothermic reaction at a given temperature
the momenta of final mesons in the center-of-mass frame equal zero.
It is shown by Eqs. (12) and (13) that the absolute values of the momenta of
initial and final mesons ($\mid \vec {P} \mid$ and $\mid 
\vec{P}^\prime \mid$) increase as $\sqrt s$ increases.
The rise of $\mid \vec {P}^\prime \mid$ causes a rapid increase of the 
cross section close to the threshold energy. The relative momentum
$\vec {p}_{ab}$ is a linear combination of $\vec {P}$ and
$\vec {P}^\prime$. Thus, $\mid \vec {p}_{ab} \mid$ increases with increasing 
$\sqrt s$, while $\psi_{ab} (\vec {p}_{ab})$ is reduced by the increase
of $\mid \vec {p}_{ab} \mid$. The wave function $\psi_{ab}
(\vec {p}_{ab})$ and the absolute value of the
transition amplitude $\mid \mathcal {M}_{\rm fi} \mid$ thus decrease with 
increasing $\sqrt s$. This decreases the cross section with increasing
$\sqrt s$. The rising $\mid \vec {P}^\prime \mid$ and falling
$\psi_{ab} (\vec {p}_{ab})$ produce a peak in the cross section near the 
threshold energy. At a higher temperature the constituents $a$ and $b$ 
have larger support of relative motion in coordinate space and so 
$\psi_{ab} (\vec {p}_{ab})$ gets narrower in momentum space. This
results in a cross section that decreases faster from the peak
and forms a more narrow peak. The exception is that the width of the 
peak of a reaction at $T/T_c=0.95$ is equal to or slightly larger than 
the one of the same reaction at $T/T_c=0.9$.

Cross sections for the seven reactions in Figs. 3-9 indicate that peak cross
sections increase from $T/T_c=0$ to 0.65 but decrease from $T/T_c=0.9$ to
0.95. While the temperature increases from zero, the potential plateau at large
distances appears and moves lower. The Schr\"odinger equation with the 
potential produces increasing meson radii. Peak cross sections increase from
$T/T_c=0$ to 0.65 as the radii of initial mesons increase.
While temperature goes to a higher value, bound states of mesons become 
looser due to weaker confinement. At a temperature near $T_c$ the bound 
states are very loose. Even though mesons are easily broken in the reaction
$A(q_{1}\overline{q}_{1})+B(q_{2}\overline{q}_{2})\to
C(q_{1}\overline{q}_{2})+D(q_{2}\overline{q}_{1})$ as $T/T_c \to 1$,
it is more difficult to combine final quarks and antiquarks into mesons
through quark rearrangement. Hence, peak cross sections decrease from 
$T/T_c=0.9$ to 0.95. On the one hand larger radii of initial mesons
cause larger cross sections for the meson-meson nonresonant reactions, 
on the other hand looser bound states of final mesons lead to smaller 
cross sections. When temperature increases, both factors generate 
rising or falling peak cross sections.

We may divide the seven reactions into three classes. The first class 
consists of the three reactions $\pi \pi \to \rho \rho$ for $I=2,$ 
$KK \to K^\ast K^\ast $ for $I=1$ and $\pi K \to \rho K^\ast$ for 
$I=3/2$ which have spin 0 for the two initial mesons and spin 1 for 
the two final mesons. The two maxima of the peak cross sections for 
the reactions are located at $T/T_c=0.9$ and 0.75, respectively.
The second class is just the reaction $\pi K^* \to \rho K$ for $I=3/2$ 
which has spin 0 for an initial meson and a final meson and spin 1 for 
the others. The maximum of the peak cross section is located at 
$T/T_c=0.85$. The third class is comprised of the three reactions
$KK^* \to K^*K^*$ for $I=1,$ $\pi K^* \to \rho K^*$ for $I=3/2$ and 
$\rho K \to \rho K^\ast$ for $I=3/2$ which have spin 0 for an initial 
meson and spin 1 for the others. The maximum of the peak cross 
sections for these reactions is located at $T/T_c=0.65$ or 0.75. In 
the first or third class the meson-flavor dependence of masses 
distinguishes the peak values of the three reactions at a given 
temperature. In the first class the difference of the total mass of 
the initial mesons and the total mass of the final mesons is larger 
than the ones in the other classes. When $T/T_c$ changes from 0.6 to 
0.99, the mass difference of the first class is reduced faster 
than the other two classes. The quicker reduction causes a more rapid 
increase of the peak value so that another maximum of the peak cross 
sections appears at $T/T_c=0.9$ in addition to the maximum at $T/T_c=0.75$.

The potential in Eq. (5) results from the interpolation between the potential 
of perturbative QCD at $r<0.01$ fm and
the lattice gauge results at large distances. The interpolation 
is in fact this procedure of adjusting the parameters $A$ and $E$.
We now begin to examine the sensitivity of mass splittings at $T=0$ GeV, 
temperature-dependent meson masses and cross sections to the interpolation
procedure by adopting two new sets of $A$ and $E$. The first set (named
Set I) is obtained by
reducing $A$ in Eq. (6) by 5\% and increasing $E$ in Eq. (7) by 10\% 
while the second set (named Set II) is obtained
by increasing $A$ in Eq. (6) by 5\% and reducing $E$ in Eq. (7) by 10\%. 
The two new sets give good fits to the lattice gauge results. Mass splittings 
at $T=0$ GeV and meson masses at five temperatures resulted from
the two sets are listed in Tables 4 and 
5, respectively. The change of the mass splittings or of the meson masses 
from Eqs. (6) and (7) to Set I or Set II is very small. The largest
change in meson mass is 2.88\% at $T=0.65T_c$. But such a change only 
leads to negligible changes in cross section. For example, the peak cross 
section obtained from Set I for $KK \to K^\ast K^\ast$ for $I=1$ at 
$T/T_c=0.65$ is
1.2735 mb in comparison to the value 1.2686 mb obtained from Eqs. (6) and (7),
and the peak cross section obtained from Set II for $\pi K \to \rho K^\ast$ 
for $I=3/2$ at
$T/T_c=0.65$ is 0.4787 mb very close to the value 0.4930 mb resulted from Eqs.
(6) and (7). Eventually, we understand that the mass splittings at $T=0$ GeV, 
the temperature-dependent masses and the 
cross sections are not sensitive to the
interpolation procedure for the construction of the potential given in Eq. (5).

In Figs. 3-9 we plot the average of the cross section in the prior form and 
the one in the post form. To see the uncertainty in this prescription, as an
example, we draw in Fig. 10 cross sections obtained in the prior form and in 
the post form for the reaction $\pi \pi \to \rho \rho$ for $I=2$. The solid 
curves in Fig. 3 are between the dashed curves corresponding to the prior form
and the dotted curves corresponding to the post form. At a given temperature
the dashed curve and the dotted curve almost overlap at the center-of-mass
energies very close to or far away from the threshold energy. The post-prior
discrepancy can
be clearly marked by the difference of peak cross sections obtained in the two
forms. We thus list the peak cross sections in Tables 6-8. The three tables 
are enough to display the discrepancy. Denote the peak 
cross section obtained in the prior (post) form by 
$\sigma^{\rm prior}_{\rm max}$ ($\sigma^{\rm post}_{\rm max}$). To indicate
the discrepancy, we define
\begin{equation}
\chi = \frac {\sigma^{\rm prior}_{\rm max} - \sigma^{\rm post}_{\rm max}}
{\sigma^{\rm prior}_{\rm max} + \sigma^{\rm post}_{\rm max}}
\end{equation}
Since $\sigma^{\rm prior}_{\rm max} > 0$  and 
$\sigma^{\rm post}_{\rm max} > 0$, 
the values of $\chi$ are between -1 and
1. The smaller the absolute value of $\chi$, the smaller the discrepancy. If 
$\chi$ is positive, $\sigma^{\rm prior}_{\rm max} > 
\sigma^{\rm post}_{\rm max}$; otherwise, $\sigma^{\rm prior}_{\rm max} \leq
\sigma^{\rm post}_{\rm max}$. The quantity $\chi$ is also presented in Tables 
6-8. From Fig. 10 we note that the peak cross sections obtained in the two 
forms at a given temperature may not correspond to the same center-of-mass
energy.

\vspace{0.5cm}
\leftline{\bf 5. Summary}
\vspace{0.5cm}

We have given the temperature-dependent central-spin-independent potential
that interpolates between the perturbative-QCD potential with loop corrections
at short distances and the potential data offered by lattice gauge calculations
at large distances. From the potential we obtain: (1) experimental mass
splittings of the ground-state mesons with the same isospin when the masses
of up, down and strange quarks are determined; (2) meson masses that
decrease from $T/T_c=0.6$ to 0.99;
(3) the wave function of quark-antiquark relative motion
that occupies a larger volume at a higher temperature. In the quark-interchange
mechanism we have obtained cross sections for seven nonresonant reactions 
$\pi \pi \to \rho \rho$ for $I=2$, $KK \to K^\ast K^\ast$ for $I=1$,
$KK^\ast \to K^\ast K^\ast$ for $I=1$, $\pi K \to \rho K^\ast$ for 
$I=3/2$, $\pi K^\ast \to \rho K^\ast$ for $I=3/2$,
$\rho K \to \rho K^\ast$ for $I=3/2$ and $\pi K^\ast \to \rho K$ 
for $I=3/2$. The temperature dependence of the cross sections is 
determined by the temperature dependence of the potential, the 
quark-antiquark wave function and the meson masses. Peak cross sections 
are affected by three factors: larger sizes of initial mesons at a 
higher temperature give larger peak cross sections, looser bound states of 
final mesons at a higher temperature lead to smaller peak cross sections, 
and a smaller total-mass difference of the initial mesons and the final 
mesons yields larger peak cross sections. The numerical cross sections are 
parametrized for future studies.

\vspace{0.5cm}
\leftline{\bf Acknowledgements}
\vspace{0.5cm}
This work was supported in part by the
National Natural Science Foundation of China
under Grant No. 10675079 and in part by Shanghai Leading Academic Discipline
Project (project number S30105). We thank H. J. Weber for a careful reading
of the manuscript.

\newpage

\begin{table}[htbp]
\centering \caption{Vacuum masses of $\pi$, $\rho$, $K$ and $K^\ast$ and their 
spin-averaged masses.}
\label{massvalue}
\begin{tabular*}{16.5cm}{@{\extracolsep{\fill}}ccccccc}
  \hline
  &
  $\overline {m}_{\pi \rho}$ (GeV) & $m_\pi$ (GeV) & $m_\rho$ (GeV) & 
  $\overline {m}_{K K^\ast}$ (GeV) & $m_K$ (GeV) & $m_{K^\ast}$ (GeV)  \\
  \hline
  \rm{model} & 0.7112 & 0.2620 & 0.8609 & 0.8339 & 0.5465 & 0.9298 \\
  \rm{experiment} & 0.6124 & 0.1396 & 0.7700 & 0.7929 & 0.4957 & 0.8920 \\
  \hline
\end{tabular*}
\end{table}

\newpage

\begin{table}[htbp]
\centering \caption{Values of parameters in the parametrization given in Eq.
(24).}
\label{paravalu1}
\begin{tabular*}{16.5cm}{@{\extracolsep{\fill}}cccccccc}
  \hline
  reaction & $T/T_c$ & $a_1~({\rm mb})$ & $b_1~({\rm GeV})$ & $c_1$
                     & $a_2~({\rm mb})$ & $b_2~({\rm GeV})$ & $c_2$  \\
  \hline
  $I=2~\pi \pi \to \rho \rho$
  & 0.65 & 0.52 & 0.13 & 2.47 & 0.07 & 0.04 & 0.42\\
  $I=2~\pi \pi \to \rho \rho$
  & 0.75 & 0.86 & 0.13 & 3.11 & 0.35 & 0.04 & 0.71\\
  $I=2~\pi \pi \to \rho \rho$
  & 0.85 & 0.65 & 0.0075 & 0.38 & 0.97 & 0.12 & 1.93\\
  $I=2~\pi \pi \to \rho \rho$
  & 0.9 & 1.96 & 0.009 & 0.5 & 0.43 & 0.15 & 1.58\\
  $I=2~\pi \pi \to \rho \rho$
  & 0.95 & 1.03 & 0.017 & 0.47 & 0.18 & 0.28 & 3.55\\
  $I=1~KK \to K^\ast K^\ast$
  & 0.65 & 0.85 & 0.11 & 0.89 & 0.4 & 0.1 & 0.44 \\
  $I=1~KK \to K^\ast K^\ast$
  & 0.75 & 1.42 & 0.09 & 1.14 & 0.46 & 0.15 & 0.39 \\
  $I=1~KK \to K^\ast K^\ast$
  & 0.85 & 1 & 0.06 & 0.76 & 0.47 & 0.08 & 0.22 \\
  $I=1~KK \to K^\ast K^\ast$
  & 0.9 & 1.3 & 0.005 & 0.63 & 0.72 & 0.04 & 0.23 \\
  $I=1~KK \to K^\ast K^\ast$
  & 0.95 & 1.56 & 0.006 & 0.48 & 0.3 & 0.065 & 0.31 \\
  $I=1~K K^\ast \to K^\ast K^\ast$
  & 0.65 & 1.71 & 0.09 & 0.58 & 0.66 & 0.1 & 2.64 \\
  $I=1~K K^\ast \to K^\ast K^\ast$
  & 0.75 & 1.63 & 0.09 & 0.71 & 0.83 & 0.07 & 1.61 \\
  $I=1~K K^\ast \to K^\ast K^\ast$
  & 0.85 & 0.88 & 0.038 & 0.57 & 0.49 & 0.1 & 0.49 \\
  $I=1~K K^\ast \to K^\ast K^\ast$
  & 0.9 & 0.33 & 0.006 & 1.17 & 0.57 & 0.035 & 0.21 \\
  $I=1~K K^\ast \to K^\ast K^\ast$
  & 0.95 & 0.51 & 0.006 & 0.47 & 0.24 & 0.05 & 0.28 \\
  \hline
\end{tabular*}
\end{table}

\newpage

\begin{table}[htbp]
\centering \caption{Values of parameters in the parametrization given in Eq.
(24).}
\label{paravalu2}
\begin{tabular*}{16.5cm}{@{\extracolsep{\fill}}cccccccc}
  \hline
  reaction & $T/T_c$ & $a_1~({\rm mb})$ & $b_1~({\rm GeV})$ & $c_1$
                     & $a_2~({\rm mb})$ & $b_2~({\rm GeV})$ & $c_2$  \\
  \hline
  $I=\frac {3}{2}~\pi K \to \rho K^\ast$
  & 0.65 & 0.3 & 0.01 & 0.89 & 0.48 & 0.16 & 1.45 \\
  $I=\frac {3}{2}~\pi K \to \rho K^\ast$
  & 0.75 & 0.44 & 0.14 & 1.43 & 0.22 & 0.1 & 0.47 \\
  $I=\frac {3}{2}~\pi K \to \rho K^\ast$
  & 0.85 & 0.09 & 0.006 & 0.001 & 0.41 & 0.093 & 0.96 \\
  $I=\frac {3}{2}~\pi K \to \rho K^\ast$
  & 0.9 & 0.76 & 0.007 & 0.65 & 0.24 & 0.05 & 0.16 \\
  $I=\frac {3}{2}~\pi K \to \rho K^\ast$
  & 0.95 & 0.55 & 0.012 & 0.59 & 0.09 & 0.03 & 0.15 \\
  $I=\frac {3}{2}~\pi K^\ast \to \rho K^\ast$
  & 0.65 & 0.72 & 0.13 & 1.29 & 0.06 & 0.06 & 0.19 \\
  $I=\frac {3}{2}~\pi K^\ast \to \rho K^\ast$
  & 0.75 & 0.74 & 0.11 & 0.93 & 0.05 & 0.04 & 0.42 \\
  $I=\frac {3}{2}~\pi K^\ast \to \rho K^\ast$
  & 0.85 & 0.22 & 0.084 & 1.95 & 0.21 & 0.08 & 0.26 \\
  $I=\frac {3}{2}~\pi K^\ast \to \rho K^\ast$
  & 0.9 & 0.25 & 0.007 & 0.63 & 0.19 & 0.05 & 0.24 \\
  $I=\frac {3}{2}~\pi K^\ast \to \rho K^\ast$
  & 0.95 & 0.16 & 0.01 & 0.56 & 0.08 & 0.07 & 0.24 \\
  $I=\frac {3}{2}~\rho K \to \rho K^\ast $
  & 0.65 & 0.7 & 0.001 & 2.65 & 1.1 & 0.12 & 1.2 \\
  $I=\frac {3}{2}~\rho K \to \rho K^\ast $
  & 0.75 & 0.69 & 0.11 & 0.81 & 0.36 & 0.09 & 2.1 \\
  $I=\frac {3}{2}~\rho K \to \rho K^\ast $
  & 0.85 & 0.1 & 0.11 & 0.12 & 0.36 & 0.07 & 0.84 \\
  $I=\frac {3}{2}~\rho K \to \rho K^\ast $
  & 0.9 & 0.27 & 0.007 & 0.53 & 0.21 & 0.05 & 0.29 \\
  $I=\frac {3}{2}~\rho K \to \rho K^\ast $
  & 0.95 & 0.09 & 0.045 & 0.14 & 0.09 & 0.03 & 4.54 \\
  $I=\frac {3}{2}~\pi K^\ast \to \rho K$
  & 0.65 & 1.68 & 0.012 & 0.53 & 0.46 & 0.06 & 0.38 \\
  $I=\frac {3}{2}~\pi K^\ast \to \rho K$
  & 0.75 & 2.12 & 0.01 & 0.76 & 0.66 & 0.01 & 0.09 \\
  $I=\frac {3}{2}~\pi K^\ast \to \rho K$
  & 0.85 & 1.82 & 0.004 & 0.48 & 1.31 & 0.01 & 0.46 \\
  $I=\frac {3}{2}~\pi K^\ast \to \rho K$
  & 0.9 & 1.6 & 0.005 & 0.42 & 0.29 & 0.03 & 0.68 \\
  $I=\frac {3}{2}~\pi K^\ast \to \rho K$
  & 0.95 & 0.58 & 0.006 & 0.4 & 0.2 & 0.04 & 1.18 \\
  \hline
\end{tabular*}
\end{table}

\newpage

\begin{table}[htbp]
\centering \caption{Mass splittings in units of GeV at zero temperature.}
\label{masssplit}
\begin{tabular*}{15cm}{@{\extracolsep{\fill}}cccc}
  \hline
  &
  $m_\rho - m_\pi$ & $m_{K^*} - m_K$ & 
  $\frac {1}{3}m_\omega + \frac {2}{3}m_\phi - m_\eta$  \\
  \hline
  Eqs. (6) and (7) & 0.5989 & 0.3833 & 0.3622 \\
  Set I            & 0.5890 & 0.3770 & 0.3530 \\
  Set II           & 0.6090 & 0.3898 & 0.3666 \\
  \hline
\end{tabular*}
\end{table}

\newpage

\begin{table}[htbp]
\centering \caption{Masses of $\pi$, $\rho$, $K$ and $K^\ast$ in units of GeV
at various temperatures.}
\label{mass}
\begin{tabular*}{15cm}{@{\extracolsep{\fill}}cccccc}
  \hline
  & $T/T_c$ & $m_\pi$ & $m_\rho$ & $m_K$ & $m_{K^\ast}$  \\
  \hline
  Eqs. (6) and (7)
  & 0.65 & 0.4042 & 0.5776 & 0.6161 & 0.7271 \\
  Eqs. (6) and (7)
  & 0.75 & 0.3839 & 0.4710 & 0.5789 & 0.6347 \\
  Eqs. (6) and (7)
  & 0.85 & 0.2958 & 0.3183 & 0.4776 & 0.4921 \\
  Eqs. (6) and (7)
  & 0.9  & 0.2105 & 0.2221 & 0.3902 & 0.3976 \\
  Eqs. (6) and (7)
  & 0.95 & 0.1021 & 0.1095 & 0.2809 & 0.2857 \\
  Set I
  & 0.65 & 0.3926 & 0.5761 & 0.6063 & 0.7237 \\
  Set I
  & 0.75 & 0.3775 & 0.4708 & 0.5736 & 0.6334 \\
  Set I
  & 0.85 & 0.2956 & 0.3182 & 0.4774 & 0.4919 \\
  Set I
  & 0.9  & 0.2110 & 0.2221 & 0.3905 & 0.3976 \\
  Set I
  & 0.95 & 0.1026 & 0.1095 & 0.2813 & 0.2858 \\
  Set II
  & 0.65 & 0.4118 & 0.5787 & 0.6162 & 0.7230 \\
  Set II
  & 0.75 & 0.3873 & 0.4713 & 0.5791 & 0.6328 \\
  Set II
  & 0.85 & 0.2952 & 0.3184 & 0.4772 & 0.4920 \\
  Set II
  & 0.9  & 0.2098 & 0.2221 & 0.3899 & 0.3978 \\
  Set II
  & 0.95 & 0.1014 & 0.1095 & 0.2807 & 0.2859 \\
  \hline
\end{tabular*}
\end{table}

\newpage

\begin{table}[htbp]
\centering \caption{Post-prior discrepancy}
\label{peakcs1}
\begin{tabular*}{15cm}{@{\extracolsep{\fill}}ccccc}
  \hline
  reaction & $T/T_c$ & $\sigma_{\rm max}^{\rm prior}$ (mb) & 
  $\sigma_{\rm max}^{\rm post}$ (mb) & $\chi$  \\
  \hline
  $I=2~\pi \pi \to \rho \rho$
  & 0 & 0.5383 & 0.4443 & 0.0957 \\
  $I=2~\pi \pi \to \rho \rho$
  & 0.65 & 0.6208 & 0.5157 & 0.0925 \\
  $I=2~\pi \pi \to \rho \rho$
  & 0.75 & 1.2168 & 0.9064 & 0.1462 \\
  $I=2~\pi \pi \to \rho \rho$
  & 0.85 & 1.2073 & 0.9010 & 0.1453 \\
  $I=2~\pi \pi \to \rho \rho$
  & 0.9 & 1.6551 & 2.2077 & -0.1431 \\
  $I=2~\pi \pi \to \rho \rho$
  & 0.95 & 0.8296 & 1.2089 & -0.1861 \\
  $I=1~KK \to K^\ast K^\ast$
  & 0 & 0.5386 & 0.7257 & -0.1480 \\
  $I=1~KK \to K^\ast K^\ast$
  & 0.65 & 1.2308 & 1.3510 & -0.0466 \\
  $I=1~KK \to K^\ast K^\ast$
  & 0.75 & 1.8543 & 1.8601 & -0.0016 \\
  $I=1~KK \to K^\ast K^\ast$
  & 0.85 & 1.7125 & 1.3213 & 0.1289 \\
  $I=1~KK \to K^\ast K^\ast$
  & 0.9 & 1.5737 & 2.2874 & -0.1848 \\
  $I=1~KK \to K^\ast K^\ast$
  & 0.95 & 1.5242 & 2.0874 & -0.1559 \\
  $I=1~K K^\ast \to K^\ast K^\ast$
  & 0 & 0.9475 & 0.9425 & 0.0026 \\
  $I=1~K K^\ast \to K^\ast K^\ast$
  & 0.65 & 2.1129 & 2.6930 & -0.1207 \\
  $I=1~K K^\ast \to K^\ast K^\ast$
  & 0.75 & 2.3603 & 2.6151 & -0.0512 \\
  $I=1~K K^\ast \to K^\ast K^\ast$
  & 0.85 & 1.4742 & 1.1657 & 0.1169 \\
  $I=1~K K^\ast \to K^\ast K^\ast$
  & 0.9 & 1.6670 & 0.0377 & 0.9558 \\
  $I=1~K K^\ast \to K^\ast K^\ast$
  & 0.95 & 1.3724 & 0.0163 & 0.9765 \\
  \hline
\end{tabular*}
\end{table}

\newpage

\begin{table}[htbp]
\centering \caption{Post-prior discrepancy}
\label{peakcs2}
\begin{tabular*}{15cm}{@{\extracolsep{\fill}}ccccc}
  \hline
  reaction & $T/T_c$ &  $\sigma_{\rm max}^{\rm prior}$ (mb) &
  $\sigma_{\rm max}^{\rm post}$ (mb) & $\chi$  \\
  \hline
  $I=\frac {3}{2}~\pi K \to \rho K^\ast$
  & 0 & 0.3430 & 0.3615 & -0.0263  \\
  $I=\frac {3}{2}~\pi K \to \rho K^\ast$
  & 0.65 & 0.5111 & 0.4830 & 0.0283  \\
  $I=\frac {3}{2}~\pi K \to \rho K^\ast$
  & 0.75 & 0.7084 & 0.6268 & 0.0611  \\
  $I=\frac {3}{2}~\pi K \to \rho K^\ast$
  & 0.85 & 0.6149 & 0.4901 & 0.1129  \\
  $I=\frac {3}{2}~\pi K \to \rho K^\ast$
  & 0.9 & 0.7691 & 1.1911 & -0.2153  \\
  $I=\frac {3}{2}~\pi K \to \rho K^\ast$
  & 0.95 & 0.4953 & 0.7650 & -0.2140  \\
  $I=\frac {3}{2}~\pi K^\ast \to \rho K^\ast$
  & 0 & 0.5415 & 0.4310 & 0.1136 \\
  $I=\frac {3}{2}~\pi K^\ast \to \rho K^\ast$
  & 0.65 & 0.7573 & 0.8279 & -0.0445  \\
  $I=\frac {3}{2}~\pi K^\ast \to \rho K^\ast$
  & 0.75 & 0.8269 & 0.8191 & 0.0047  \\
  $I=\frac {3}{2}~\pi K^\ast \to \rho K^\ast$
  & 0.85 & 0.5147 & 0.4022 & 0.1227  \\
  $I=\frac {3}{2}~\pi K^\ast \to \rho K^\ast$
  & 0.9 & 0.6862 & 0.1273 & 0.6870  \\
  $I=\frac {3}{2}~\pi K^\ast \to \rho K^\ast$
  & 0.95 & 0.3985 & 0.0823 & 0.6576  \\
  $I=\frac {3}{2}~\rho K \to \rho K^\ast $
  & 0 & 0.6318 & 0.5553 & 0.0644  \\
  $I=\frac {3}{2}~\rho K \to \rho K^\ast $
  & 0.65 & 1.1125 & 1.1724 & -0.0262  \\
  $I=\frac {3}{2}~\rho K \to \rho K^\ast $
  & 0.75 & 1.1026 & 1.0686 & 0.0157  \\
  $I=\frac {3}{2}~\rho K \to \rho K^\ast $
  & 0.85 & 0.5808 & 0.4368 & 0.1415  \\
  $I=\frac {3}{2}~\rho K \to \rho K^\ast $
  & 0.9 & 0.8565 & 0.0061 & 0.9859  \\
  $I=\frac {3}{2}~\rho K \to \rho K^\ast $
  & 0.95 & 0.4294 & 0.0080 & 0.9634  \\
  \hline
\end{tabular*}
\end{table}

\newpage

\begin{table}[htbp]
\centering \caption{Post-prior discrepancy}
\label{peakcs3}
\begin{tabular*}{15cm}{@{\extracolsep{\fill}}ccccc}
  \hline
  reaction & $T/T_c$ &  $\sigma_{\rm max}^{\rm prior}$ (mb) &
  $\sigma_{\rm max}^{\rm post}$ (mb) & $\chi$  \\
  \hline
  $I=\frac {3}{2}~\pi K^\ast \to \rho K$
  & 0 & 1.4320 & 1.1445 & 0.1116  \\
  $I=\frac {3}{2}~\pi K^\ast \to \rho K$
  & 0.65 & 2.7801 & 0.7960 & 0.5548  \\
  $I=\frac {3}{2}~\pi K^\ast \to \rho K$
  & 0.75 & 2.9691 & 1.1931 & 0.4267  \\
  $I=\frac {3}{2}~\pi K^\ast \to \rho K$
  & 0.85 & 3.3879 & 2.6925 & 0.1144  \\
  $I=\frac {3}{2}~\pi K^\ast \to \rho K$
  & 0.9 & 1.8472 & 1.7048 & 0.0401  \\
  $I=\frac {3}{2}~\pi K^\ast \to \rho K$
  & 0.95 & 0.6578 & 0.6441 & 0.0105  \\
  \hline
\end{tabular*}
\end{table}

\newpage

\begin{figure}[htbp]
\centering
\includegraphics[width=15cm]{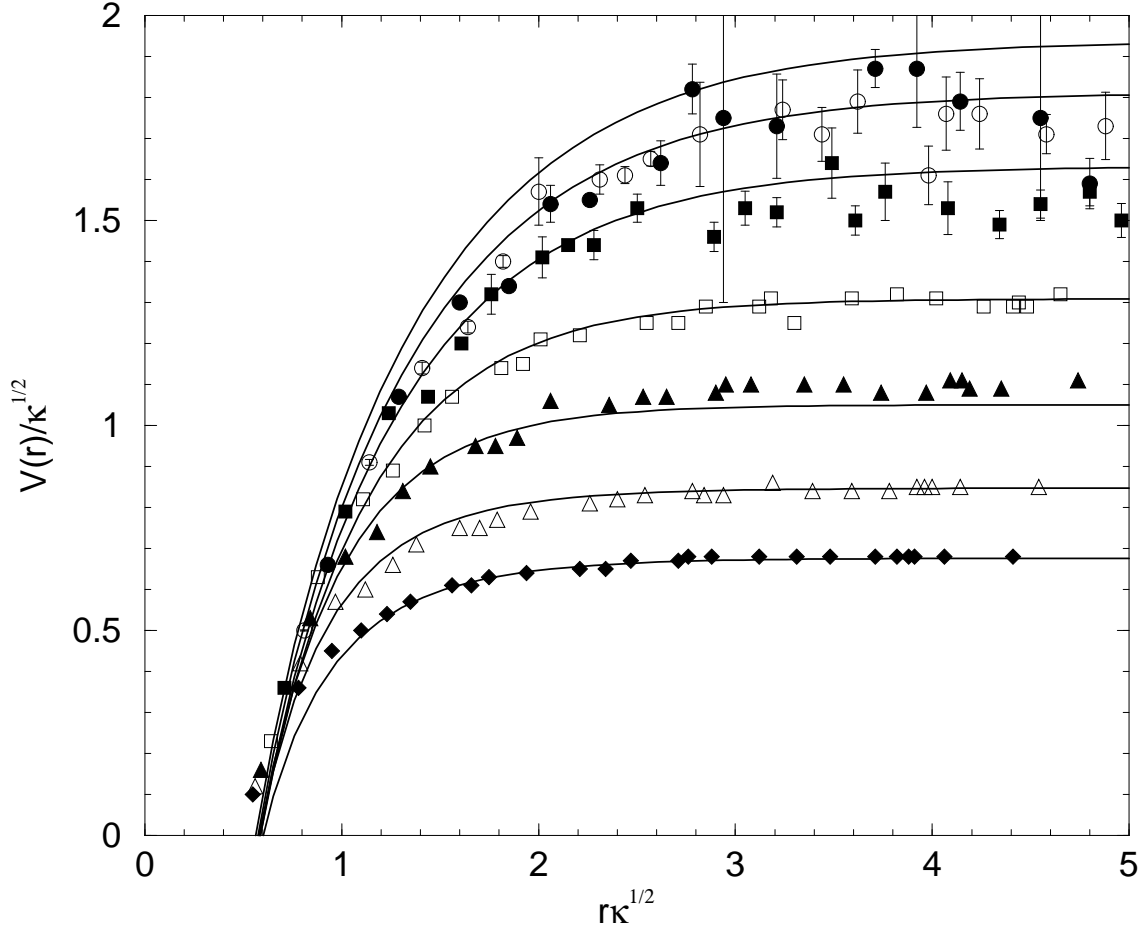}
\caption{Temperature-dependent potential. From top to bottom
the temperatures corresponding to the different data sets \cite{KLP} are
$T/T_c=0.58,~0.66,~0.74,~0.84,~0.9,~0.94,~0.97$.
Solid curves stand for the parametrization fitted to the data at these
temperatures. $\sqrt {\kappa} =2.154~{\rm fm}^{-1}=0.425~{\rm GeV}$.}
\label{fig1}
\end{figure}

\newpage

\begin{figure}[htbp]
\centering
\includegraphics[scale=0.9]{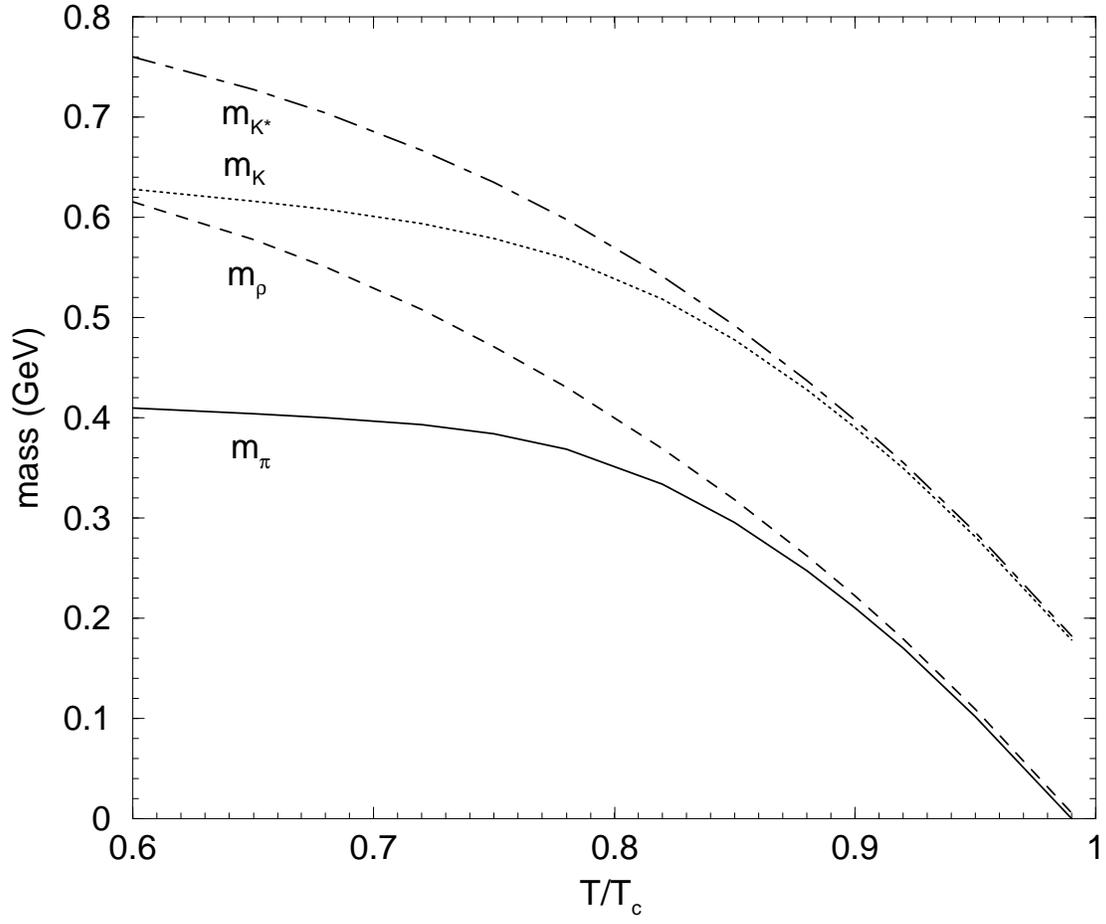}%
\caption{Meson masses as functions of $T/T_c$.}
\label{fig2}
\end{figure}

\newpage

\begin{figure}[htbp]
\centering
\includegraphics[scale=0.9]{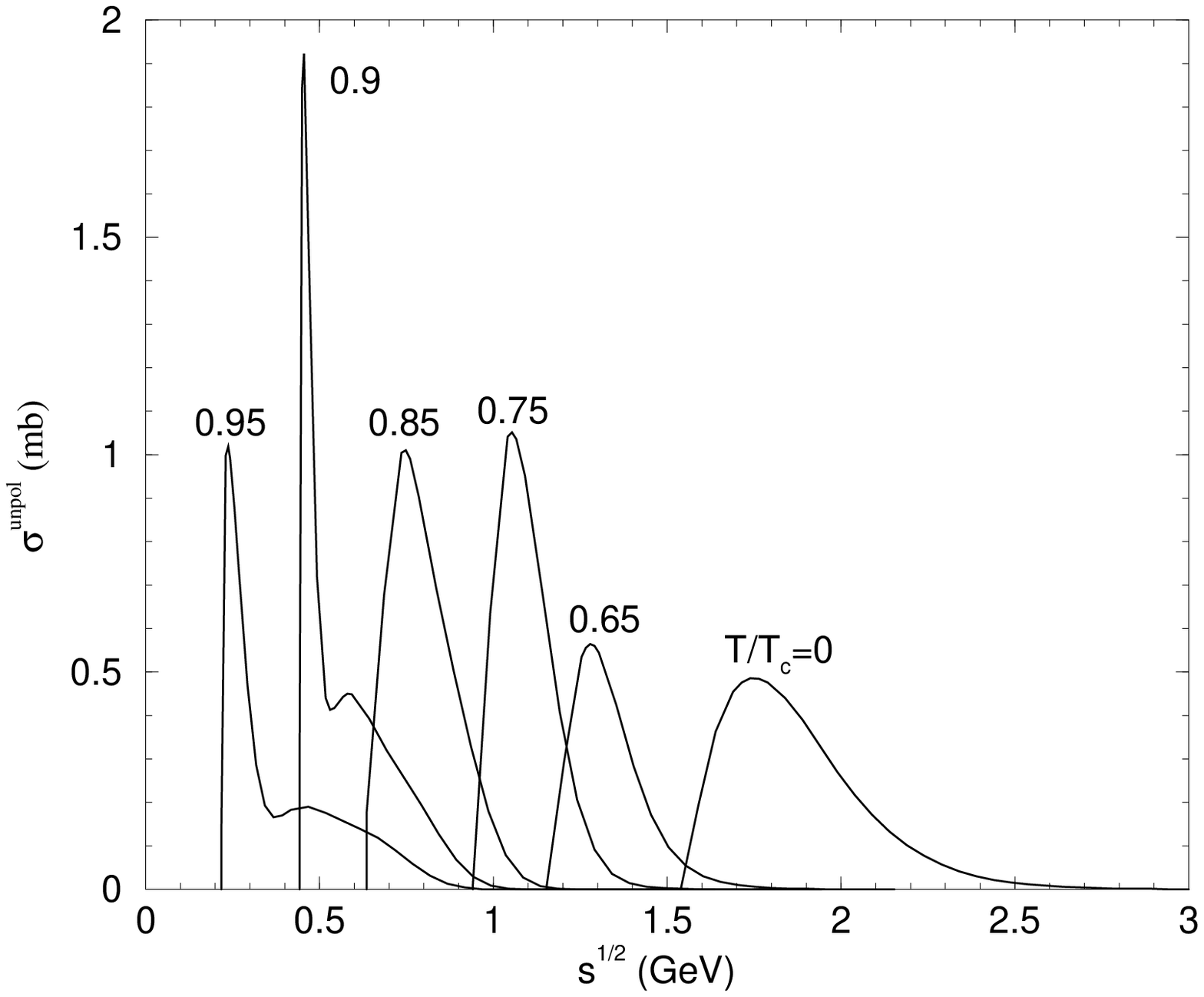}%
\caption{$\pi \pi \to \rho \rho $ cross sections for $I=2$ at various
temperatures.}
\label{fig3}
\end{figure}

\newpage

\begin{figure}[htbp]
\centering
\includegraphics[scale=0.9]{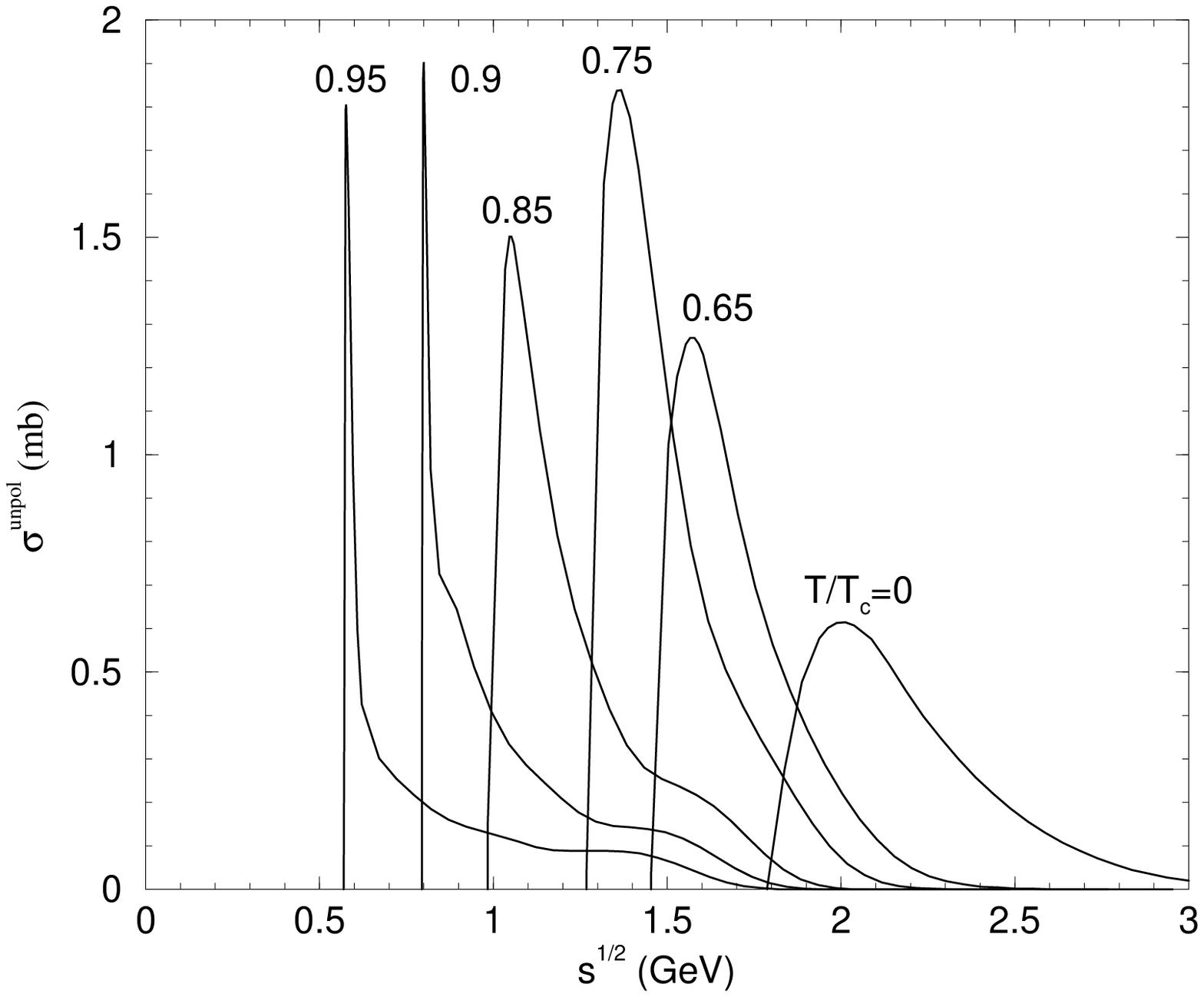}%
\caption{$K K \to K^\ast K^\ast $ cross sections for $I=1$ at various
temperatures.}
\label{fig4}
\end{figure}

\newpage

\begin{figure}[htbp]
\centering
\includegraphics[scale=0.9]{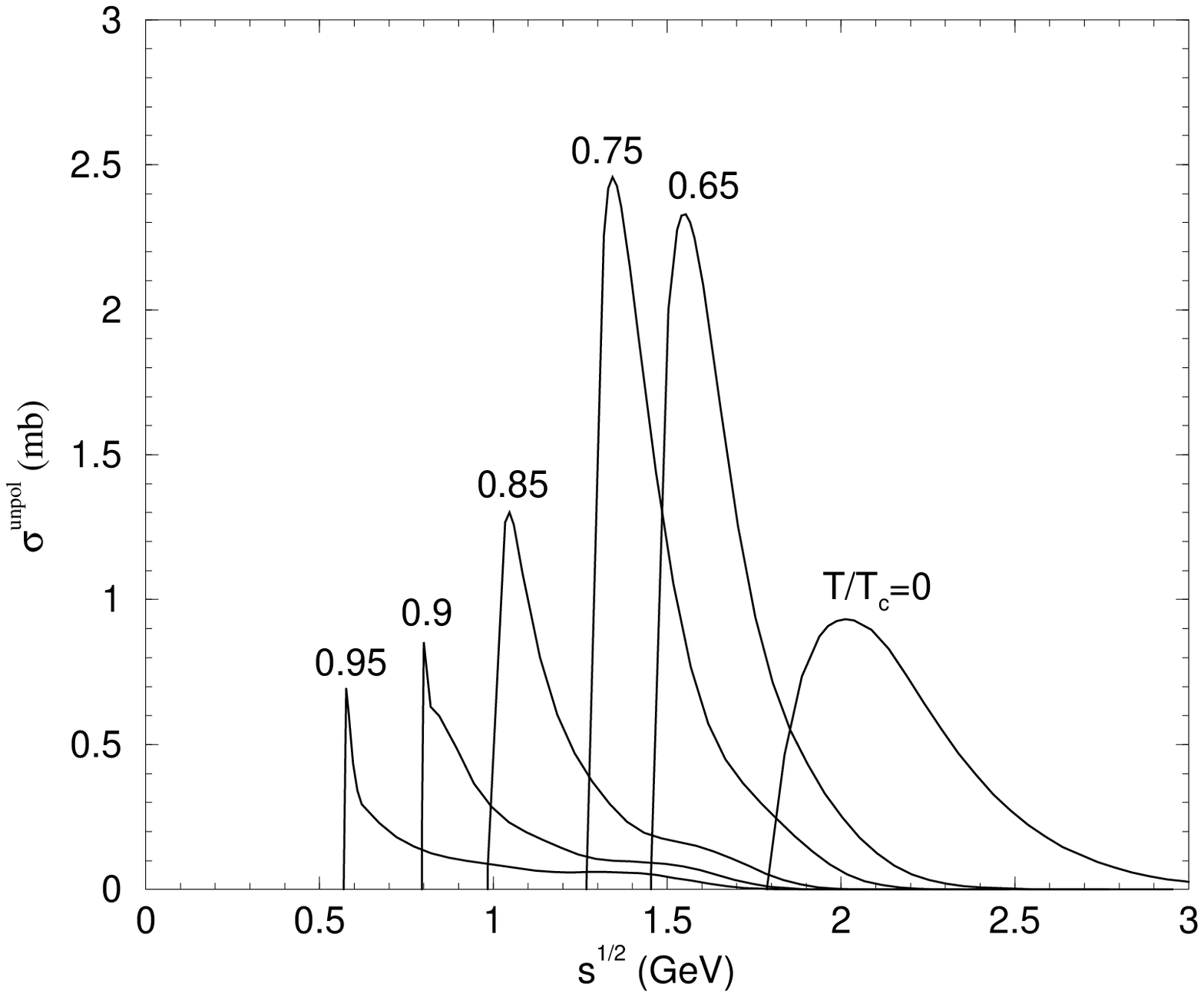}
\caption{$K K^\ast \to K^\ast K^\ast $ cross sections for $I=1$ at various
temperatures.}
\label{fig5}
\end{figure}

\newpage

\begin{figure}[htbp]
\centering
\includegraphics[scale=0.9]{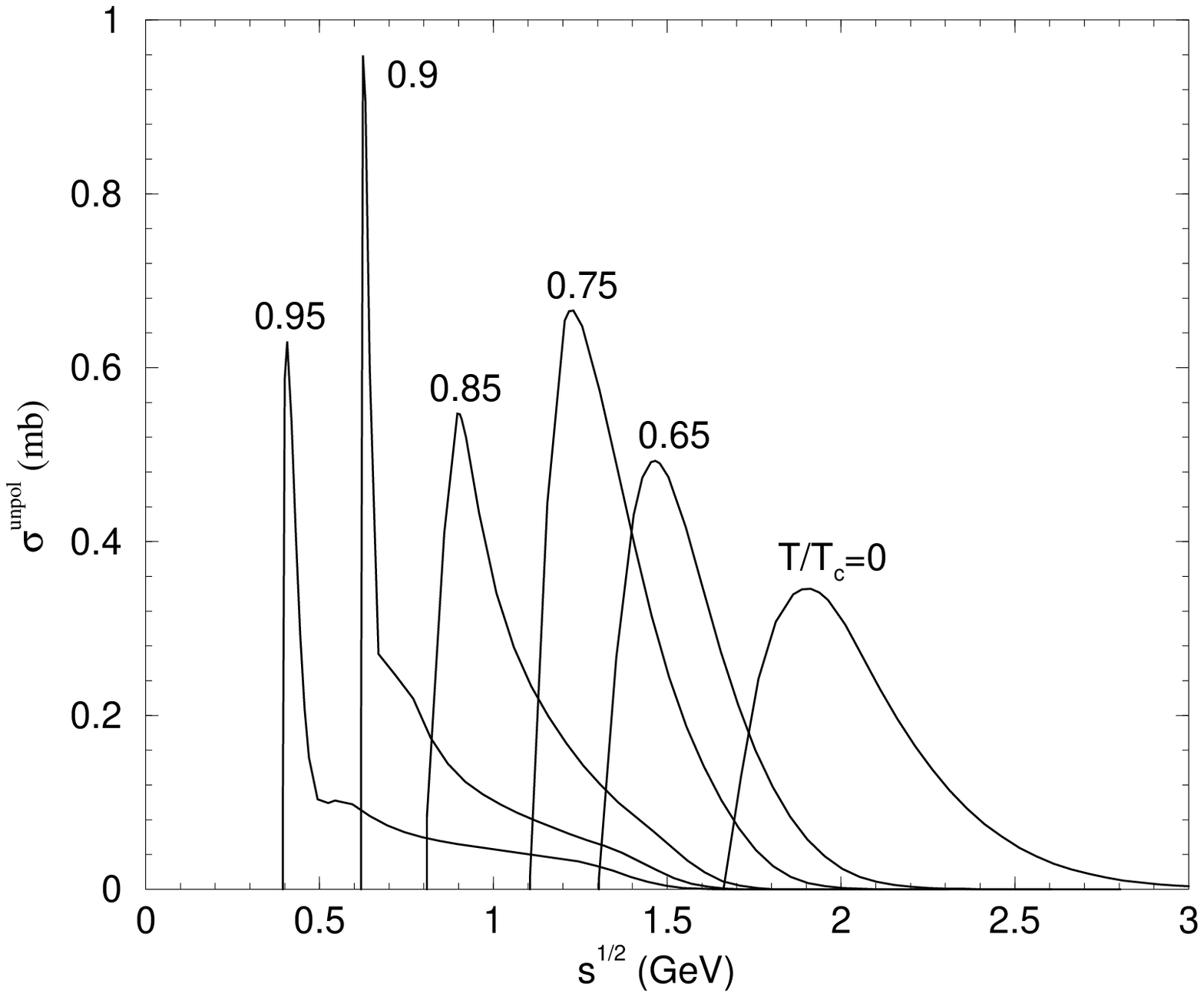}%
\caption{$\pi K \to \rho K^\ast$ cross sections for $I=3/2$ at various
temperatures.}
\label{fig6}
\end{figure}

\newpage

\begin{figure}[htbp]
\centering
\includegraphics[scale=0.9]{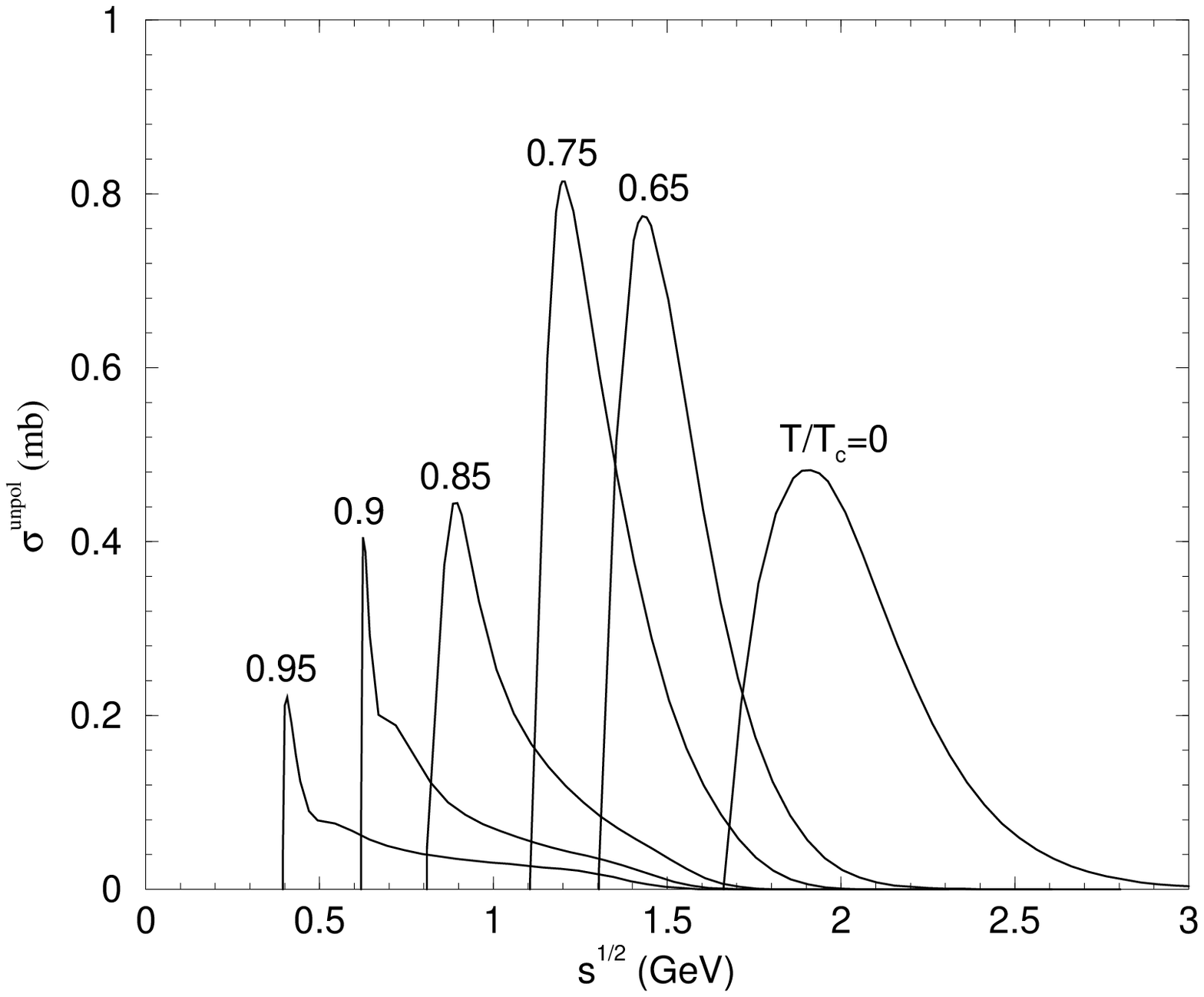}
\caption{$\pi K^\ast \to \rho K^\ast$ cross sections for $I=3/2$ at various
temperatures.}
\label{fig7}
\end{figure}

\newpage

\begin{figure}[htbp]
\centering
\includegraphics[scale=0.9]{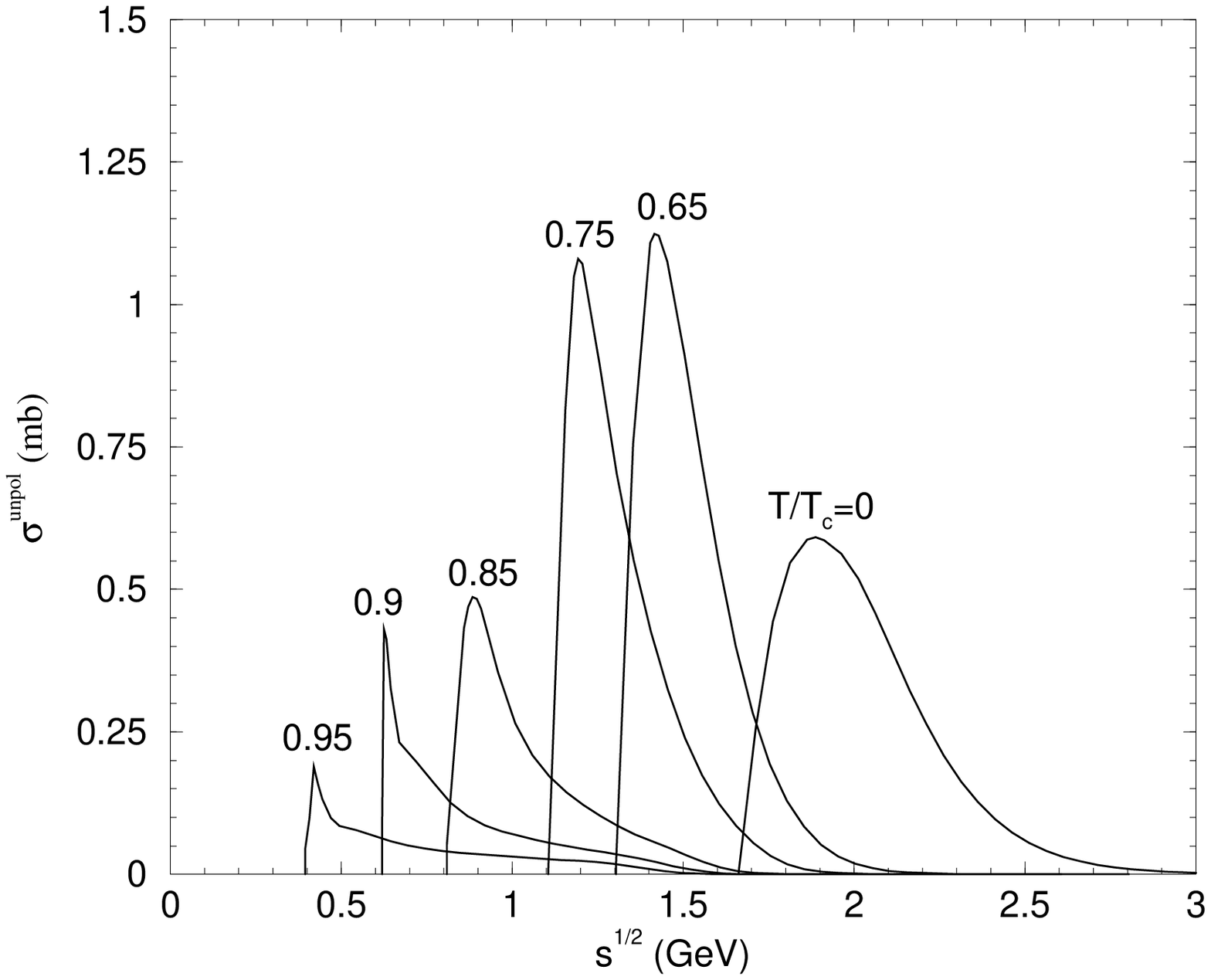}
\caption{$\rho K \to \rho K^\ast $ cross sections for $I=3/2$ at various
temperatures.}
\label{fig8}
\end{figure}

\newpage

\begin{figure}[htbp]
\centering
\includegraphics[scale=0.9]{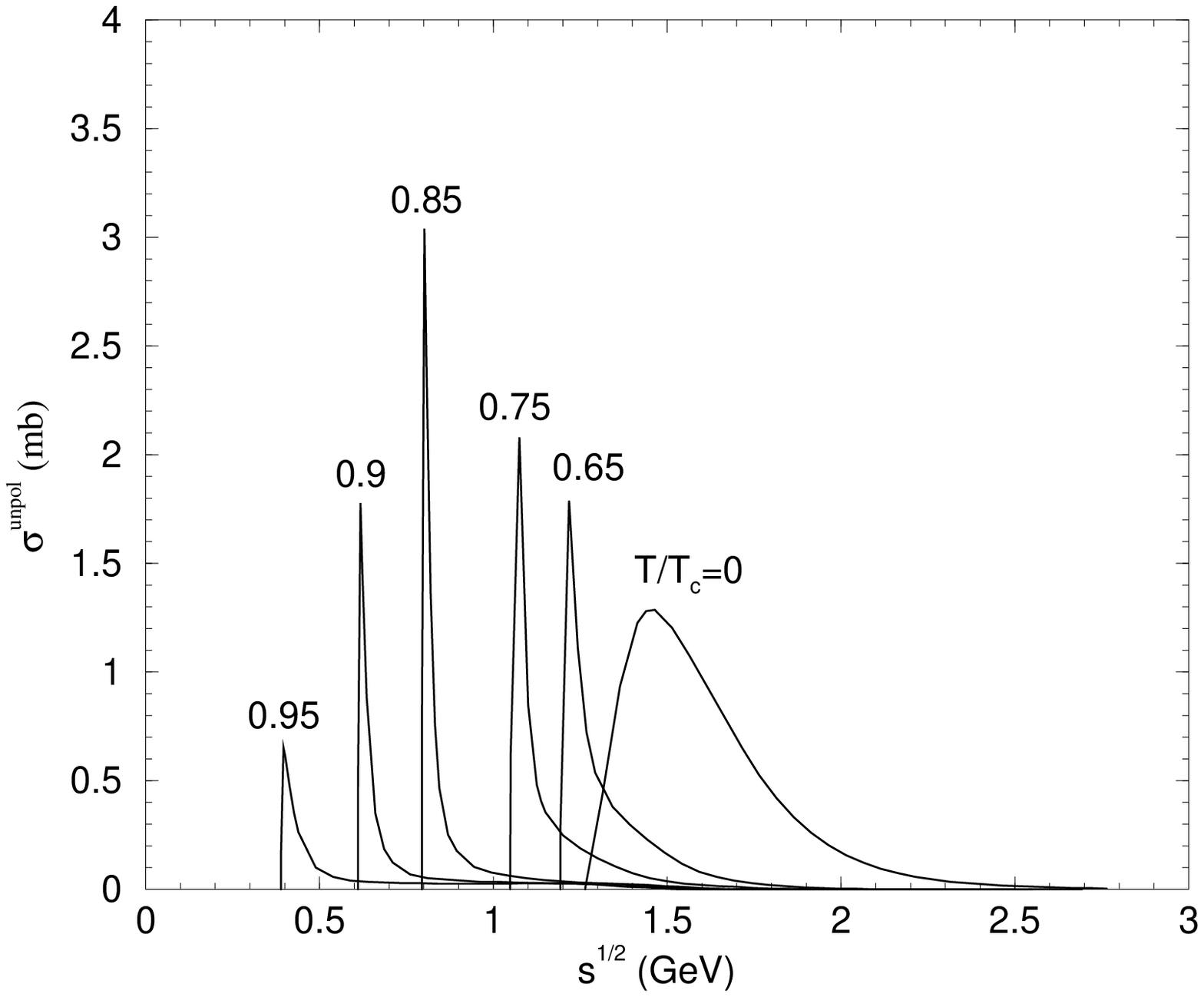}%
\caption{$\pi K^\ast \to \rho K$ cross sections for $I=3/2$ at various
temperatures.}
\label{fig9}
\end{figure}

\newpage

\begin{figure}[htbp]
\centering
\includegraphics[scale=0.9]{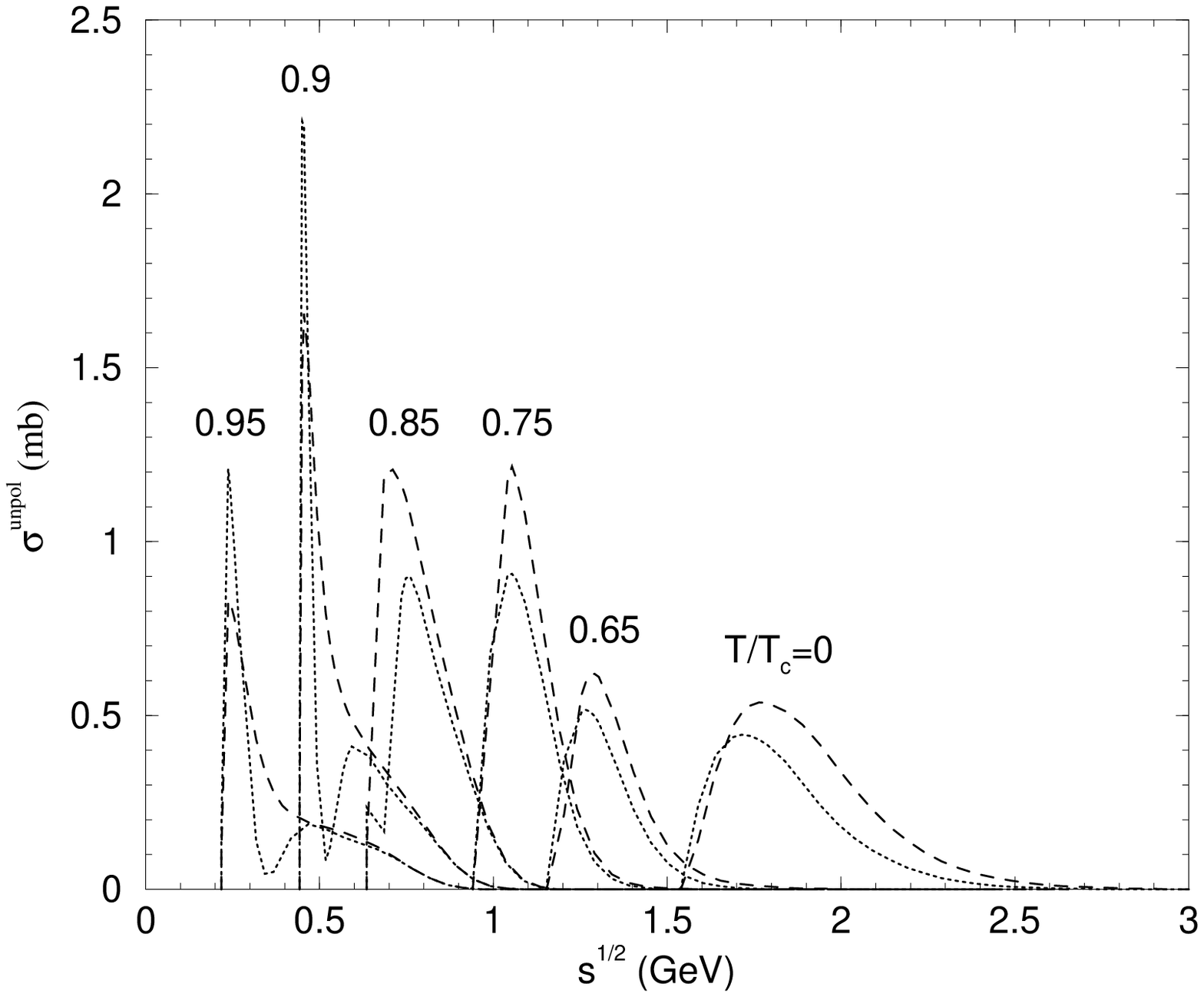}%
\caption{$\pi \pi \to \rho \rho$ cross sections for $I=2$ in the prior form
(dashed curves) and in the post form (dotted curves) at various temperatures.}
\label{fig10}
\end{figure}


\newpage
\begin{thebibliography}{99}
\addtolength{\itemsep}{-0.6 em}
\bibitem{BS92}T. Barnes, E.S. Swanson, Phys. Rev. D 46 (1992) 131.
\bibitem{Swanson}E.S. Swanson, Ann. Phys. (N.Y.) 220 (1992) 73.
\bibitem{MBQ}K. Martins, D. Blaschke, E. Quack, Phys. Rev. C 51 (1995) 2723.
\bibitem{WSB}C.-Y. Wong, E.S. Swanson, T. Barnes, Phys. Rev. C
65 (2001) 014903.
\bibitem{BSWX}T. Barnes, E.S. Swanson, C.-Y. Wong, X.-M. Xu, Phys. Rev. C 68
(2003) 014903.
\bibitem{Xu02}X.-M. Xu, Nucl. Phys. A 697 (2002) 825.
\bibitem{SK}D. Kharzeev, H. Satz, Phys. Lett. B 334 (1994) 155.
\bibitem{Peskin}M.E. Peskin, Nucl. Phys. B 156 (1979) 365.
\bibitem{BP}G. Bhanot, M.E. Peskin, Nucl. Phys. B 156 (1979) 391.
\bibitem{Duraes}F.D. Duraes, et al., Phys. Rev. C 68 (2003) 035208.
\bibitem{Mat98}S.G. Matinyan, B. M\"uller, Phys. Rev. C 58 (1998) 2994.
\bibitem{Lin00}Z. Lin, C.M. Ko, J. Phys. G 27 (2001) 617.
\bibitem{Hag99}K. Haglin, Phys. Rev. C 61 (2000) 031902.
\bibitem{Hag01}K. Haglin, C. Gale, Phys. Rev. C 63 (2001) 065201.
\bibitem{Oh01}Y. Oh, T. Song, S.H. Lee, Phys. Rev. C 63 (2001) 034901.
\bibitem{adler}S.S. Adler, et al., PHENIX Collaboration, Phys. Rev. C 69 (2004)
034909.
\bibitem{arsene}I. Arsene, et al., BRAHMS Collaboration, Nucl. Phys. A 757
(2005) 1.
\bibitem{back}B.B. Back, et al., PHOBOS Collaboration, Nucl. Phys. A 757 (2005)
28.
\bibitem{adams1}J. Adams, et al., STAR Collaboration, Nucl. Phys. A 757 (2005)
102.
\bibitem{adcox}K. Adcox, et al., PHENIX Collaboration, Nucl. Phys. A 757 (2005)
184.
\bibitem{adams2}J. Adams, et al., STAR Collaboration, Phys. Rev. Lett. 92
(2004) 092301.
\bibitem{bearden1}I.G. Bearden, et al., BRAHMS Collaboration, Phys. Rev. Lett.
90 (2003) 102301.
\bibitem{bearden2}I.G. Bearden, et al., BRAHMS Collaboration, Phys. Rev. Lett.
94 (2005) 162301.
\bibitem{LX}Y.-Q. Li, X.-M. Xu, Nucl. Phys. A 794 (2007) 210.
\bibitem{KLP}F. Karsch, E. Laermann, A. Peikert, Nucl. Phys. B 605 (2001) 579.
\bibitem{BDPK}I. Bender, H.G. Dosch, H.J. Pirner, H.G. Kruse, Nucl. Phys. A
414 (1984) 359.
\bibitem{DHKK}M. D$\ddot {\rm o}$ring, K. H$\ddot {\rm u}$bner, O. Kaczmarek,
F. Karsch, Phys. Rev. D 75 (2007) 054504.
\bibitem{GHK}S. Gupta, K. H$\ddot {\rm u}$bner, O. Kaczmarek,
Phys. Rev. D 77 (2008) 034503.
\bibitem{Michael}C. Michael, arXiv:hep-ph/9809211.
\bibitem{BaPi}G.S. Bali, A. Pineda, Phys. Rev. D 69 (2004) 094001.
\bibitem{Sjostrand}T. Sj$\ddot {\rm o}$strand, Comput. Phys. Commun. 39 (1986) 
347.
\bibitem{Miyazawa}H. Miyazawa, Phys. Rev. D 20 (1979) 2953.
\bibitem{GS}V. Goloviznin, H. Satz, Yad. Fiz. 60N3 (1997) 523.
\bibitem{Satz06hep}H. Satz, hep-ph/0602245.
\bibitem{KKPZ}O. Kaczmarek, F. Karsch, P. Petreczky, F. Zantow, Phys. Lett.
B 543 (2002) 41.
\bibitem{ZKKP}F. Zantow, O. Kaczmarek, F. Karsch, P. Petreczky, 
hep-lat/0301015.
\bibitem{Satz09}H. Satz, arXiv:0812.3829.
\bibitem{KS}L. Kluberg, H. Satz, arXiv:0901.3831.
\bibitem{DPS1}S. Digal, P. Petreczky, H. Satz, Phys. Lett. B 514 (2001) 57.
\bibitem{SZ}E. Shuryak, I. Zahed, Phys. Rev. D 70 (2004) 054507.
\bibitem{Wong05}C.-Y. Wong, Phys. Rev. C 72 (2005) 034906.
\bibitem{Wong}C.-Y. Wong, hep-ph/0509088.
\bibitem{Wong07}C.-Y. Wong, Phys. Rev. C 76 (2007) 014902.
\bibitem{Alberico}W.M. Alberico, A. Beraudo, A. De Pace, A. Molinari, Phys. 
Rev. D 72 (2005) 114011.
\bibitem{Satz06}H. Satz, J. Phys. G 32 (2006) R25.
\bibitem{BT}W. Buchm\"{u}ller, S.-H.H. Tye, Phys. Rev. D 24 (1981) 132.
\bibitem{Wong02}C.-Y. Wong, Phys. Rev. C 65 (2002) 034902.
\bibitem{DGG}A. De R$\acute{\rm u}$jula, H. Georgi, S.L. Glashow, Phys. Rev. D 
12 (1975) 147.
\bibitem{IS1}N. Isgur, G. Karl, Phys. Rev. D 18 (1978) 4187.
\bibitem{IS2}N. Isgur, G. Karl, Phys. Rev. D 19 (1979) 2653.
\bibitem{IS3}N. Isgur, G. Karl, Phys. Rev. D 20 (1979) 1191.
\bibitem{GI}S. Godfrey, N. Isgur, Phys. Rev. D 32 (1985) 189.
\bibitem{CI}S. Capstick, N. Isgur, Phys. Rev. D 34 (1986) 2809.
\bibitem{Chra}Z.V. Chraplyvy, Phys. Rev. 91 (1953) 388.
\bibitem{BD}J.D. Bjorken, S.D. Drell, Relativistic Quantum Mechanics,
McGraw-Hill, New York, 1964.
\bibitem{Weinberg}S. Weinberg, The Quantum Theory of Fields, Vol. II, Cambridge
University Press, Cambridge, 1996.
\bibitem{Pokorski}S. Pokorski, Gauge Field Theories, Cambridge University 
Press, Cambridge, 2000.
\bibitem{FGH}J.F. Donoghue, E. Golowich, B.R. Holstein, Dynamics of the 
Standard Model, Cambridge University Press, Cambridge, 1992.
\bibitem{GL1}J. Gasser, H. Leutwyler, Ann. Phys. 158 (1984) 142.
\bibitem{GL2}J. Gasser, H. Leutwyler, Nucl. Phys. B 250 (1985) 465.
\bibitem{BCEGS}J. Bijnens, G. Colangelo, G. Ecker, J. Gasser, M.E. Sainio, 
Nucl. Phys. B 508 (1997) 263.
\bibitem{CGL}G. Colangelo, J. Gasser, H. Leutwyler, Nucl. Phys. B 603 (2001) 
125.
\bibitem{MM}N.F. Mott, H.S.W. Massey, The Theory of Atomic Collisions,
Clarendon Press, Oxford, 1965.
\bibitem{BBS}T. Barnes, N. Black, E.S. Swanson, Phys. Rev. C 63 (2001) 025204.
\bibitem{WC}C.-Y. Wong, H.W. Crater, Phys. Rev. C 63 (2001) 044907.
\bibitem{Col71} E. Colton, et al., Phys. Rev. D 3 (1971) 2028.
\bibitem{Dur73} N.B. Durusoy, et al., Phys. Lett. B 45 (1973) 517.
\bibitem{Hoo77} W. Hoogland, et al., Nucl. Phys. B 126 (1977) 109.
\bibitem{Los74} M.J. Losty, et al., Nucl. Phys. B 69 (1974) 185.

\end{thebibliography}
\end{document}